\documentclass[useAMS,usenatbib]{mn2e}
\usepackage{graphicx,amssymb, txfonts,pstricks}

\def\pb{Pa$\beta$}
\def\br{Br$\gamma$}

\def\hb{H$\beta$}
\def\feii{[Fe\,{\sc ii}]}
\def\pii{[P\,{\sc ii}]}
\def\oiii{[O\,{\sc iii}]}

\def\h2{H$_2$}
\def\p1{Paper~I}

\def\kms {$\rm km\,s^{-1}$}

\title[Feeding vs. Feedback in NGC\,5929]{Feeding versus feedback in AGN from near-infrared Integral Field Spectroscopy X: NGC\,5929}
\author[Riffel, Storchi-Bergmann \& Riffel]{Rogemar A. Riffel$^{1,2}$\thanks{E-mail:
rogemar@ufsm.br}, Thaisa Storchi-Bergmann$^{2}$ and Rog\'erio Riffel$^{2}$
\\
$^{1}$ Departamento de F\'\i sica, Centro de Ci\^encias Naturais e Exatas, Universidade Federal de Santa Maria, 97105-900, Santa Maria, RS, Brazil \\ 
$^{2}$ Departamento de Astronomia, Instituto de F\'\i sica, Universidade Federal do Rio Grande do Sul, CP 15051, 91501-970, Porto Alegre, RS, Brazil }

\begin{document}

\date{Accepted 2014 December 15. Received 2014 December 14; in original form 2014 December 13}

\pagerange{\pageref{firstpage}--\pageref{lastpage}} \pubyear{2014}

\maketitle

\label{firstpage}

\begin{abstract}

We present near-infrared emission-line flux distributions, excitation and kinematics, as well as stellar kinematics, of the inner $520\times520~{\rm pc^2}$ of the Seyfert 2 galaxy NGC\,5929. The observations were performed with the Gemini's Near-Infrared Integral Field Spectrograph (NIFS) at a spatial resolution of $\sim20~$pc and spectral resolution of $40~$\kms\ in the  J- and K$_{\rm l}$-bands.
The flux distributions of \h2, \feii, \pii, and H recombination lines are extended over most of the field of view, with the highest intensity levels observed along $PA=60/240^\circ$, and well correlated with the radio emission. The \h2\ and \feii\ line emission are originated in thermal processes, mainly due to heating of the gas by X-rays from the central Active Galactic Nucleus (AGN). Contribution of shocks due to the radio jet is observed at locations co-spatial with the radio hotspots at 0\farcs50 northeast and 0\farcs60 southwest of the nucleus, as evidenced by the emission-line ratio and gas kinematics. 
The stellar kinematics shows rotation with an amplitude at 250pc from the nucleus of $\sim$200 \kms\ after corrected for the inferred inclination of 18.3$^\circ$. The stellar velocity dispersion obtained from the integrated K-band spectrum is $\sigma_*=133\pm8$~\kms, which implying on a mass for the supermassive black hole of  $M_\bullet=5.2^{1.6}_{-1.2}\times10^7~{\rm M_\odot}$, using the  $M_\bullet-\sigma_*$ relation.
The gas kinematics present three components: (1) gas in the plane of the galaxy in counter-rotation relative to the stars; (2) an outflow perpendicular to the radio jet that seems to be due to an equatorial AGN outflow; (3) turbulence of the gas observed in association with the radio hot spots, supporting an interaction of the radio jet with the gas of the disk. We estimated the mass of ionized and warm molecular gas  of $\sim1.3\times10^6\,{\rm M_\odot}$
  and $\sim470~{\rm M_\odot}$, respectively. 

\end{abstract}


\section{Introduction} \label{intro}

Detailed mapping of the gas distribution, excitation and kinematics around Active Galactic Nuclei (AGN)
is fundamental to the understanding of the physics behind the AGN feeding and feedback processes. Near-IR integral field spectroscopy (IFS) 
on 8-10 m telescopes of nearby galaxies has become a powerful tool to better understand 
these processes, as it provides a two-dimensional coverage with spatial resolution of a 
few to tens of parsecs at a spectral resolution that allows to resolve the gas kinematics and permitting isolating inflows and
outflows in the central kiloparsec, where the relevant processes occur \citep[e.g.][]{n4051,ms09,davies09,sb10,schartmann10,mrk1066-kin,mrk79,davies14,paperI}.

Some of the main results obtained from IFS of the inner kpc of active galaxies are: (I) ionized and molecular gas have distinct
flux distributions and kinematics. The near-IR line-emission at these scales is originated by the heating and ionization of the gas by the AGN radiation and shocks produced by radio jets  \citep[e.g.][]{eso428,mrk1066-exc,mrk79}. The molecular \h2\ emitting gas is usually more restricted to the plane of galaxies, with kinematics dominated by rotation and inflows in the disk in most cases. The ionized gas emission traces a more disturbed medium, usually associated to outflows from the AGN, but frequently showing also a rotation component from the disk of the galaxy \citep[e.g.][]{mrk1066-exc,mrk1066-kin,mrk1157,sb10,iserlohe13,mazzalay14}. (II) Outflows are seen within hollow cones or from compact structures with mass outflow rates in the range $10^{-2} - 10^{1}~{\rm M_\odot yr^{-1}}$ for low-luminosity Seyfert and Low-Ionization Nuclear Emission-line Regions (LINERs) galaxies \citep[e.g.][]{n7582,schonel14} and $10^{2} - 10^{3}~{\rm M_\odot yr^{-1}}$ for high-luminosity Seyferts \citep[e.g.][]{mcelroy14}. The outflows are observed with velocities from 200~\kms\ to 800~\kms in ionized gas \citep{sb14a} and $\sim$150~\kms\ in \h2\, when present \citep{davies14}. (III) Inflows are observed in \h2\ \citep[e.g.][]{ms06,mrk79,mazzalay14} in Seyfert galaxies and in low-ionization gas in LINERs \citep{n1097,n6951,m81,n2110-opt,n7213} with mass inflow rates in the range $10^{-1} - 10^{1}~{\rm M_\odot yr^{-1}}$ \citep{sb14a,sb14b}. (IV) The stellar kinematics in Seyfert galaxies reveal cold nuclear structures composed of young stars, usually associated with a significant gas reservoir \citep{hicks13,mrk1066-pop,mrk1157-pop,sb12}.

NGC~5929 is a spiral galaxy with a Seyfert 2 nucleus located at a distance of 35.9~Mpc\footnote{as quoted in NASA/IPAC EXTRAGALACTIC DATABASE --http://ned.ipac.caltech.edu/}. NGC~5929 has a companion, NGC~5930 at 20$^{\prime\prime}$ northeast from it \citep{page52}.  
It presents a well defined bi-polar radio jet oriented along the position angle PA$\approx$60$^\circ$, showing a triple structure with two bright hot spots, one located at 0\farcs5 north-east from the nucleus and the other at 0\farcs6 south-west from it. The third and fainter radio structure is observed at the nucleus of the galaxy \citep{ulvestad84,wilson89,su96}. The orientation of the major axis of the large scale disk is  PA$\approx$45$^\circ$\citep[e.g.][]{schmitt97}. 

The gas kinematics of NGC~5929 has been studied for at least 3 decades. Using long slit spectroscopy along the orientation of the radio jet, \citet{keel85} concluded that the ionized gas kinematics is consisten with rotation in the galaxy disk. \citet{whittle86} obtained long-slit spectra along PA=60$^\circ$ and PA=$-$30$^\circ$ covering the H$\beta$ and {O\,{\sc iii}] emission lines and found double-component emission-line profiles along PA=$-$30$^\circ$. They  suggested that the origin of these profiles was  the superposition of the two components observed along the PA=60$^\circ$ (one in blueshift to the north-east and one in redshift to the south-west)  and unresolved by their observations, at a seeing of 1\farcs6.  \citet{rosario10} report the detection of shocked gas associated to the north-east radio hot spot by comparing optical long slit spectra obtained with the Hubble Space Telescope Imaging Spectrograph (HST/STIS)  with radio images. These authors found that the low ionization gas (traced by H$\beta$ emission) shows broader emission-line profiles than the high ionization gas (traced by [O\,{\sc iii}]$\lambda5007$ emission line) at locations near the radio hot spot and suggested that this broadening is due to the interaction of the radio jet with the NLR. Evidence of shocks is also observed at the location of the south-western radio hot spot, for which \citet{ferruit99} showed that shock models with velocities of $\approx$300\,km\,s$^{1}$ are able to reproduce the optical and UV emission-line ratios. The low spectral and spatial resolutions and/or small spatial coverage of the data have not allowed the studies above to reveal the complete scenario for the gas kinematics in the nuclear region of NGC\,5929, what we have been able to do now with our NIFS observations.

In \citet[][hereafter called Paper I]{paperI} we have reported already the discovery of a peculiar structure along PA=$-$30/150$^\circ$ in this galaxy from these observations, where the emission lines present double components. This structure was interpreted as being due to the interaction of the acquired gas from the companion with an ``equatorial outflow'' from the accretion disk around the central supermassive black hole (SMBH). In the present paper, we present a more complete study of the overall gas kinematics, as well as of the gas excitation and stellar kinematics of the inner $520\times520~{\rm pc^2}$ of NGC~5929.

NGC\,5929 represents the tenth galaxy of a series observed by our group AGNIFS ({\it AGN Integral Field Spectroscopy}) in which we present the emission-line flux distributions and kinematics as well as the stellar kinematics of the inner few hundreds of parsec of nearby active galaxies. The previous papers of this series have presented similar studies of the following galaxies: 
I - ESO\,428-G14 \citep{eso428};
II- NGC\,4051 \citep{n4051};
III - NGC\,7582 \citep{n7582};
IV - NGC\,4151 \citep{sb09,sb10};
V - Mrk\,1066 \citep{mrk1066-exc,mrk1066-kin,mrk1066-pop};
VI - Mrk\,1157 \citep{mrk1157,mrk1157-pop};
VII - Mrk\,79 \citep{mrk79};
VIII - NGC\,1068 \citep{n1068-exc,sb12,n1068-kin};
IX - Mrk\,766 \citep{schonel14}.

 This paper is organized as follows: In Section~2 we present the description of the observations and data reduction. In Sec.~3 we present the results 
for the continuum and line emission, including line-ratio maps as well as the gas and stellar kinematics. The discussion of the results  
is presented in Sec.~4 and the conclusions are presented in Sec.~5.

\section{Observations and Data Reduction}

NGC\,5929 was observed using the Gemini Near-infrared Integral Field Spectrograph \citep[NIFS --][]{mcgregor03} on the Gemini North telescope operating with the adaptive optics module ALTAIR on the nights of March 16, May 24 and June 16, 2011 under the observation Programme GN-2011A-Q-43.  The observations covered the J and K$_{\rm l}$ spectral bands, resulting in a wavelength coverage from 1.14$\mu$m to 1.36$\mu$m  and 2.10$\mu$m to 2.54$\mu$m, respectively. The total on source exposure time for each band was 6000\,s and the observations were splited into 10 individual on source exposures plus 5 sky exposures for each band.

The data reduction followed standard procedures and was accomplished using tasks contained in the {\sc nifs.gemini}
package which is part of {\sc iraf} software. The procedures included the trimming of the images,
flat-fielding, sky subtraction, wavelength and s-distortion calibrations, remotion of the telluric absorptions and flux calibration  by
interpolating a black body function to the spectrum of the telluric standard star. 

The final data cubes contain $\sim$4500 spectra at a angular sampling of 0\farcs05$\times$0\farcs05, covering the inner 3$^{\prime \prime}\times$3$^{\prime \prime}$ ( 520$\times$520~pc$^2$) of NGC\,5929. The angular resolution is 0\farcs12 for both bands (corresponding to $\sim$20\,pc at the galaxy), as obtained from the full width at half maximum (FWHM) of the images of the telluric standard stars. From the FWHM of the emission-line profiles of the Ar lamps we obtain the spectral resolution of 1.7\,\AA\ for the J band and  3.2\,\AA\ for the K band, corresponding to velocity resolutions of $\sim$40\,km\,s$^{-1}$ for both bands.

\begin{figure*}
\centering
\includegraphics[scale=0.65]{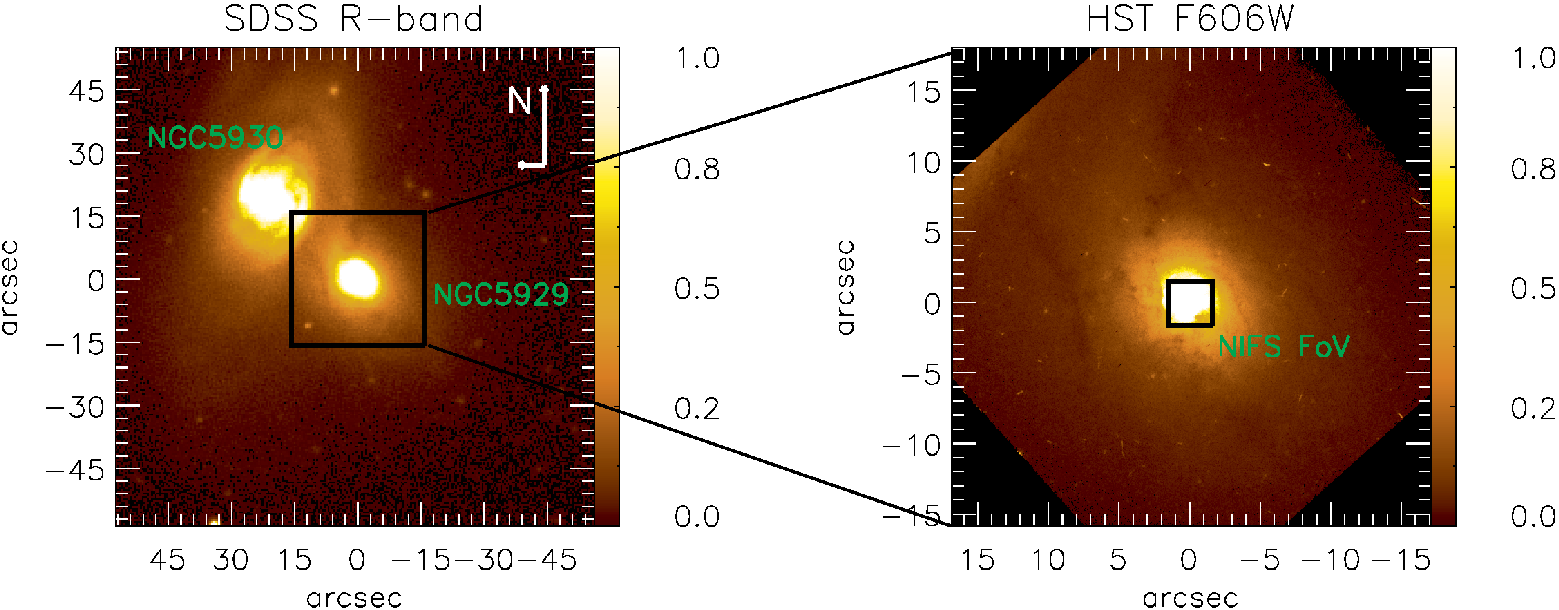}
\includegraphics[scale=0.85]{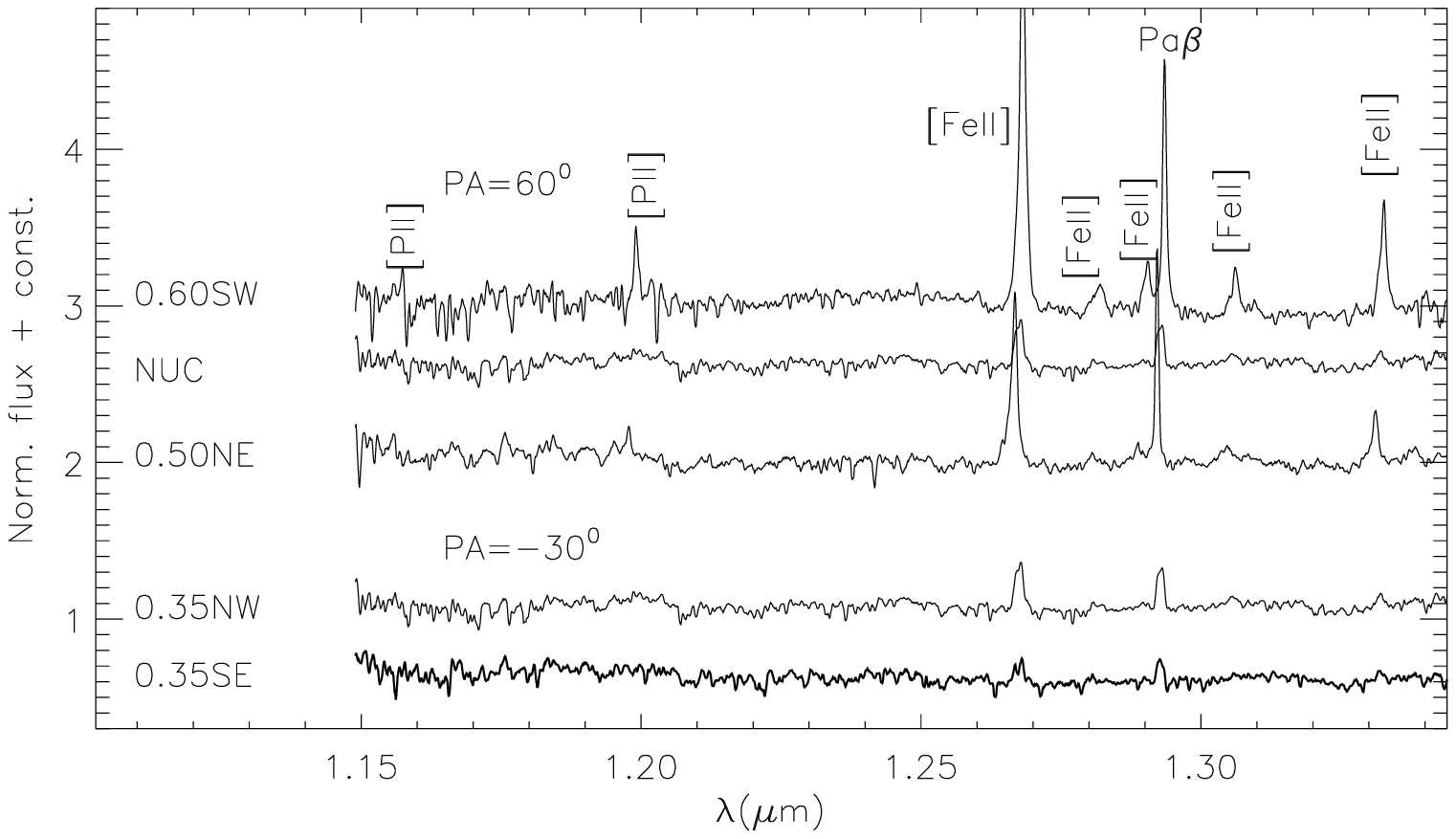} 
\includegraphics[scale=0.85]{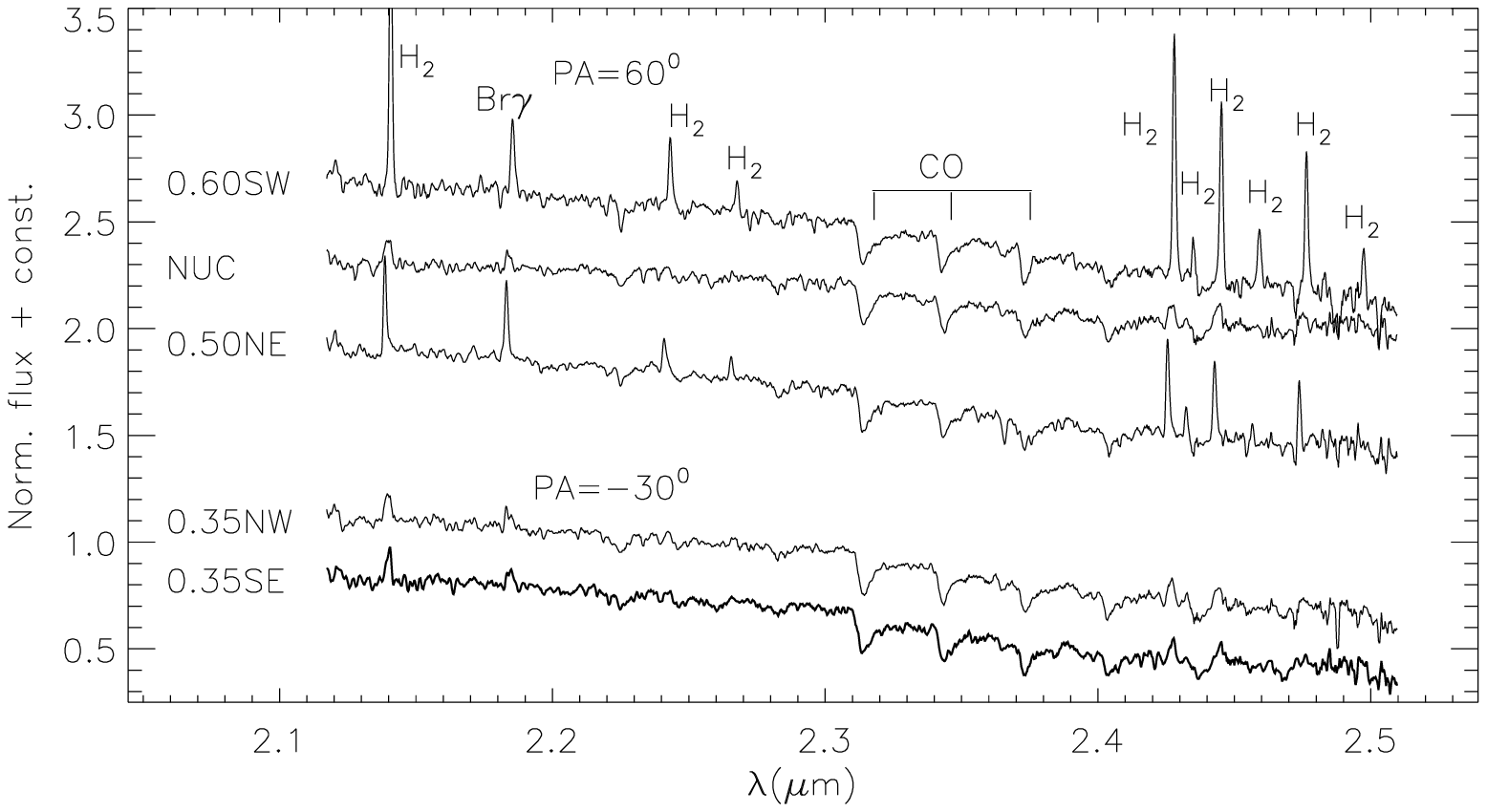}
\caption{Top left: Large scale image of NGC\,5929 and NGC\,5930 obtained in the R-band from the SDSS DR4 \citep{baillard11}. 
 Top right: Optical image of the inner 30\arcsec$\times$30\arcsec of NGC~5929 obtained with the HST WFPC2 trhought the filter F606W \citep{malkan98},
 with the NIFS field of view represented by the green box. 
Middle: NIFS J-band sample spectra for the central region of NGC\,5929, with the main emission lines identified.  Bottom: NIFS K-band sample spectra for the central region of NGC\,5929 with the main emission/absorption lines identified. The spectra shown were integrated within an aperture of 0\farcs25$\times$0\farcs25 and the positions selected correspond to the south-western radio hotspot (0\farcs60 SW of the nucleus), for the nucleus (NUC), the  position of the north-eastern radio hotspot (0\farcs50 NE) and for two positions along PA=$-30^\circ$ (0\farcs35 NW and 0\farcs35 SE), along of the peculiar outflow reported in Paper I. }
\label{large}
\end{figure*}

In the top-left panel of Fig.\,\ref{large} we present a large scale $R$-band image of NGC~5929 and its companion NGC~5930 from the Sloan Digital Sky Survey, covering 1.5\arcmin$\times$1.5\arcmin \citep{baillard11}. The top-right panel shows an optical image of NGC~5929 obtained with the Hubble Space Telescope (HST) Wide Field Planetary Camera 2 (WFPC2) through the broad-band filter F606W from \citet{malkan98}, covering the inner 30\arcsec$\times$30\arcsec. The green box represents the field of view of our NIFS data. The middle panel shows sample spectra obtained from the NIFS  J-band datacube within apertures of 0\farcs05$\times$0\farcs05 for the following positions (from top to bottom): 0\farcs6 southwest of the nucleus, corresponding to the position of the southwestern radio hotspot \citep{ulvestad89};  0\farcs5 northeast of the nucleus, corresponding to the position of the northeastern radio hotspot  \citep{ulvestad89}; the nucleus;  0\farcs35 northwest and 0\farcs35 southeast of the nucleus along $PA=-30^\circ$, perpendicular to the radio jet, where we found that the emission lines present double components \citep{paperI}.  The following emission lines are identified in the J-band spectra: [P {\sc ii}]$\lambda$1.14713\,, [P {\sc ii}]$\lambda$1.18861\,$\mu$m, [Fe\,{\sc ii}]$\lambda1.25702\,\mu$m, [Fe\,{\sc ii}]$\lambda1.27069\,\mu$m, [Fe\,{\sc ii}]$\lambda1.27912\,\mu$m, H\,{\sc i}~Pa$\beta\,\lambda$1.28216\,$\mu$m, [Fe\,{\sc ii}]$\lambda1.29462\,\mu$m and [Fe\,{\sc ii}]$\lambda1.32092\,\mu$m. The bottom panel of Fig.~\ref{large} shows the spectra IN the K-band at the same positions of those IN the J-band. We identified the H\,{\sc i}~Br$\gamma\,\lambda$2.16612\,$\mu$m emission line and the H$_2$ emission lines at  2.12183, 2.15420, 2.22344, 2.24776, 2.40847, 2.41367, 2.42180, 
2.43697 and 2.45485\,$\mu$m.  The CO absorption band-heads at $\sim$2.3~$\mu$m used to measure the stellar kinematics are also identified in the K-band spectra.

\section{Results}

\subsection{Emission-line profiles}

As discussed in Paper I, the emission lines present double components along a strip (hereafter identified as the SE-NW strip) crossing the nucleus at $PA=-30^\circ$, which have been attributed to gas outflowing from the nucleus perpendicularly to the radio jet. In Figure~\ref{profiles} we show a sample of the emission-line profiles of \feii$\lambda1.2570$, \h2$\lambda2.1218$ and \pb\ from three locations: along  the SE-NW strip  at 0\farcs35 southeast of the nucleus (top panel) and at the positions of the radio knots along $PA=60^\circ$ at 0\farcs50 north-east (middle panel) and at  0\farcs60 south-west (bottom panel) of the nucleus.  The double components, reported in Paper I, are evident at the position 0\farcs35 south-east, not only for \feii\, but also for the other emission lines, with profiles presenting similar widths. Indeed, all emission lines detected along the SE-NW strip present similar double components. At the positions of the radio hotspots or knots, the \feii\ profile is the broadest, \h2\ is the narrowest and the \pb one is in between the two.The \h2\ and \pb\ are well reproduced by a single Gaussian curve, while the \feii\ profile clearly shows the presence of more than one component. A blue wing in the \feii\ profile is observed for the position at  0\farcs50 northeast of the nucleus and at  0\farcs60 southwest of the nucleus its profile is broader than a Gaussian curve. At positions away from the strip at $PA=-30^\circ$ and distant from the radio knots the emission lines are well reproduced by a single Gaussian component, as discussed in Paper I.

\begin{figure}
\centering
\includegraphics[scale=0.5]{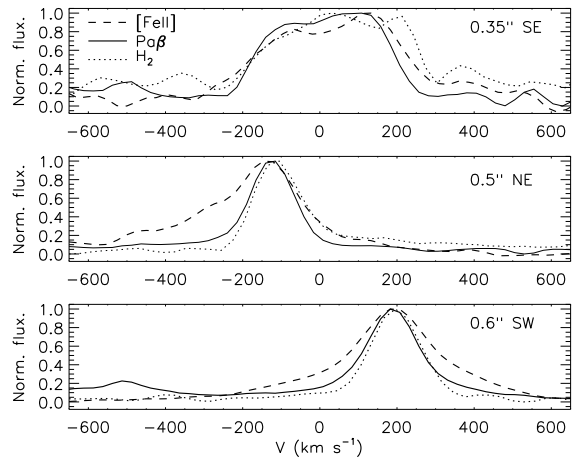}
\caption{Sample of the \feii$\lambda1.2570$, \h2$\lambda2.1218$ and \pb\ emission-line profiles at three locations: 0\farcs35 southeast of the nucleus, along the SE-NW strip at $PA=-30^\circ$ (top panel); 0\farcs50 northeast of the nucleus, at the position of a radio hot spot (middle panel) and 0\farcs60 southwest of the nucleus, at the position of the other radio hot spot (bottom panel).  The \pb\ line profile is shown as a continuous line, the \h2\ as a dotted line and the dashed line represents the \feii\ emission line profile.} 
\label{profiles}
\end{figure}

As shown in Fig.~\ref{large}, several emission lines are observed in the J- and K-band spectra of NGC~5929. In order to measure the emission line flux distributions we fitted the observed line profiles at each spatial position by Gaussian curves using the {\sc profit} routine \citet{profit}.  In  
table\,\ref{fluxes} we present the measured fluxes for the  emission lines within an aperture of 0\farcs35$\times$0\farcs35 
at five positions: the locations of the radio hotspots, at 0\farcs50 northeast and 0\farcs60 southwest of the nucleus; the nucleus; at 0\farcs35 southeast and 0\farcs35 northwest of the nucleus along the strip at $PA=-30^\circ$ (region where the equatorial outflow was observed in Paper I). Values followed by ``$^*$" were obtained by the sum of the fluxes of the two components. The \feii$\lambda1.2570$ line profile was fitted by three Gaussian curves at 0\farcs60 southeast of the nucleus (at the location of a radio hotspot), and the flux quoted in the table is the sum of the fluxes of the three components. Sample fits of the \feii\ profile are shown in Paper I.


\begin{table*}
\centering
\caption{Measured emission-line fluxes for the five positions along $PA=60^\circ$ and $PA=-30^\circ$ within 0\farcs35$\times$0\farcs35 aperture in 
units of 10$^{-16}$\,erg\,s$^{-1}$\,cm$^{-2}$.}
\vspace{0.3cm}
\begin{tabular}{l l c c c c c}
\hline

$\lambda_{vac} {\rm(\AA)}$    & ID                      & 0\farcs50NE     & 0\farcs60SW                 &  Nucleus         & 0\farcs35SE    & 0\farcs35NW    \\
\hline
11471.3  & [P {\sc ii}]\,$^1D_3-^3P_1$                  & 12.1$\pm$5.3      &  6.3$\pm$4.3            &  --              &   --             &   --    \\
11665.6   & [Fe\,{\sc ii}]\,$b^4P_{1/2}-a^2P_{3/2}$     & 21.1$\pm$4.8      &       --                &    --            &      --          &   --    \\
11863.7   & [Fe\,{\sc ii}]\,$b^4D_{1/2}-b^4F_{5/2}$     & 10.4$\pm$5.7      &       --                &     --           &        --        &   --     \\
11886.1  & [P {\sc ii}]\,$^1D_2-^3P_2$                  & 18.3$\pm$4.4      &  26.9$\pm$8.6           &  --              & 	--          &   --                \\
12570.2  & [Fe\,{\sc ii}]\,$a^4D_{7/2}-a^6D_{9/2}$      & 179.4$\pm$15.3$^*$&  299.9$\pm$18.0$^{**}$  & 110.2$\pm$8.2$^*$& 35.0$\pm$8.5$^*$ &   110.1$\pm$9.3$^*$ \\
12791.2  & [Fe\,{\sc ii}]\,$a^4D_{3/2}-a^6D_{3/2}$      & 9.9$\pm$2.3       &  4.7$\pm$1.2            &     --           &	--	    &   14.6$\pm$9.5$^*$  \\
12821.6  &  H\,{\sc i}\,Pa$\beta$                      &117.3$\pm$5.2       &  106.1$\pm$10.8         & 79.0$\pm$9.3$^*$ & 29.1$\pm$4.4$^*$ &  157.9$\pm$12.1$^*$ \\
12946.2  & [Fe\,{\sc ii}]\,$a^4D_{5/2}-a^6D_{5/2}$     & 13.4$\pm$4.4       &  5.5$\pm$3.3            &  18.6$\pm$9.5$^*$& 0.70$\pm$0.13$^*$& 17.8$\pm$14.1$^*$ 	  \\
13209.2  & [Fe\,{\sc ii}]\,$a^4D_{7/2}-a^6D_{7/2}$     & 60.9$\pm$20.4$^*$  &  69.0$\pm$7.8           &  13.6$\pm$8.2$^*$&2.21$\pm$0.26$^*$ &24.5$\pm$20.9$^*$        \\
13281.4  & [Fe\,{\sc ii}]\,$a^4D_{5/2}-a^6D_{3/2}$     & 18.8$\pm$11.2      &  --                     &  --              &	--          &  --        	   \\

21218.3  & H$_2$2\, 1-0S(1) 	  		       & 18.7$\pm$ 1.7  & 45.7$\pm$ 1.5    & 27.1$\pm$7.4$^*$  &	 16.8$\pm$4.8$^*$  &  16.5$\pm$3.0$^*$	\\
21542.0  &  H$_2$\,1-0\,S(2)  	                       & 1.9$\pm$ 0.7   & 2.1$\pm$ 0.6     & --                &	 3.8$\pm$2.9$^*$   &	-- \\
21661.2  & H\,{\sc i}\,Br$\gamma$                      &16.4$\pm$ 1.5   & 14.5$\pm$ 1.3    & 12.9$\pm$2.4$^*$  &	 10.5$\pm$3.0$^*$  &	12.5$\pm$3.5$^*$ \\
22234.4  &  H$_2$\,1-0\,S(0)  	                       & 9.4$\pm$ 0.7   & 11.4$\pm$ 0.6    & --                &	 --                &	-- \\
22477.6  &  H$_2$\,2-1\,S(1)  	                       & 4.9$\pm$0.62  & 5.2$\pm$ 1.0     &  --               &	 --                &	-- \\
24084.7  &  H$_2$\,1-0\,Q(1)   	                       &21.8$\pm$ 1.3   & 39.5$\pm$ 0.7    &25.8$\pm$6.8$^*$   &	  13.8$\pm$6.4$^*$ & 15.5$\pm$6.0$^*$ \\
24136.7  &  H$_2$\,1-0\,Q(2)   	                       & 6.3$\pm$ 2.5   &  7.9$\pm$ 1.9    &  --               &	 --                &  -- \\
24218.0  &  H$_2$\,1-0\,Q(3)   	                       & 18.0$\pm$ 2.8  & 32.3$\pm$ 2.8    & 25.0$\pm$14.4$^*$ &	  15.4$\pm$9.9$^*$ &  17.6$\pm$13.1$^*$ \\
24369.7  &  H$_2$\,1-0\,Q(4)   	                       & --             & 12.6$\pm$ 1.9    & --                &	 --                &  -- \\
24548.5  & H$_2$\,1-0\,Q(5)   	                       & 14.3$\pm$ 3.6  & 22.9$\pm$ 4.7    & --                &	 --                &  -- \\  
24755.5   & H$_2$\,1-0\,Q(6)   	                       & --             & 8.4$\pm$ 4.1     & --                &	 --                &  -- \\  
25000.7   & H$_2$\,1-0\,Q(7)   	                       & --             & 10.5$\pm$8.2     &  --               &	 --                &  --  \\  

\hline
\multicolumn{7}{l}{$^*$ Emission-line profile fitted by two Gaussian curves and the flux represents the sum of the fluxes of both components;}\\
\multicolumn{7}{l}{$^{**}$ Emission-line profile fitted by three Gaussian curves and the flux represents the sum of the fluxes of the components.}\\
\end{tabular}
\label{fluxes}
\end{table*}

\subsection{Emission-line flux distributions}

In order to obtain the flux distributions, we fitted only one Gaussian to all emission lines, for simplicity. Although two or three components reproduce better the profiles at some locations, as discussed above, the resulting error in the fluxes at these locations is lower than 5\%.

Figure~\ref{fluxmaps} shows the flux distributions for all measured
emission lines. In each panel, 
black represents masked regions where the flux values are smaller than the standard deviation of the continuum values near the emission line 
or the uncertainty in the flux is higher than 40\,\%. Light gray contours delimit the masked regions and the green contours overlaid to some panels are from the radio image of \citet{ulvestad89}. At most locations the uncertainties in flux are smaller than 15\,\%.  For illustration purpose, we have rebbined the spaxels to 1/3 of their original size and then interpolated their fluxes. As the original spaxels (0\farcs05$\times$0\farcs05) are smaller than the angular resolution of our observations, this procedure does not affect the spatial resolution of the maps significantly. The line emission is more extended along the $PA=60/240^\circ$, extending up to 1\farcs5 to both sides of the nucleus, while to the perpendicular direction ($PA=-30/150^\circ$) the emission is extended to 0\farcs7 from the nucleus. The flux distributions of all emission lines show a good correlation with the radio structures, with the two peaks of emission associated to the southwestern and northeastern radio hotspots. Some differences are observed among distinct flux distributions. While the \feii\ and \h2\ fluxes peak at the location of the southwestern  radio structure at 0\farcs6 from the nucleus, the H\,{\sc i} recombination lines present their highest fluxes at the location of the northeastern radio hotspot at 0\farcs50 from the nucleus. Another difference is that the \h2\ emission  
is less collimated and more extended in all directions, as most clearly seen in the H$_2\,\lambda$2.1218 flux distribution (that presents the highest signal-to-noise ratio among the \h2\ lines).

\begin{figure*}
\centering
\includegraphics[scale=0.88]{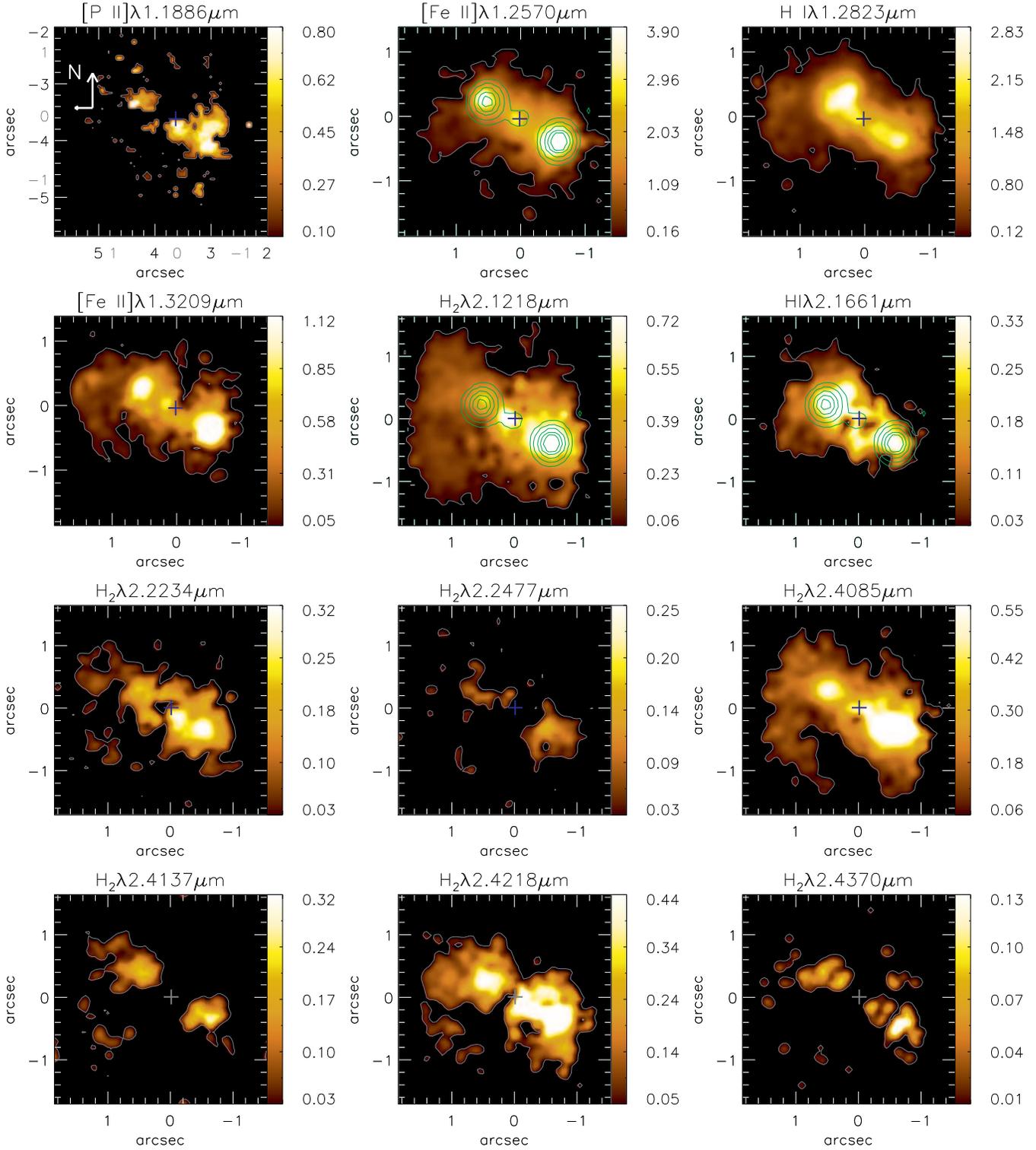}
\caption{Integrated flux maps for the emission lines identified in the top of each panel, in units of 10$^{-17}$\,erg\,s$^{-1}$\,cm$^{-2}$. The 
central cross marks the position of the continuum peak. The green contours overlaid on some maps are from the 6\,cm radio continuum image from \citet{ulvestad89}.}
\label{fluxmaps}
\end{figure*}

\subsection{Line-ratio maps}\label{sec_ratio}

\begin{figure*}
\centering
\includegraphics[scale=0.9]{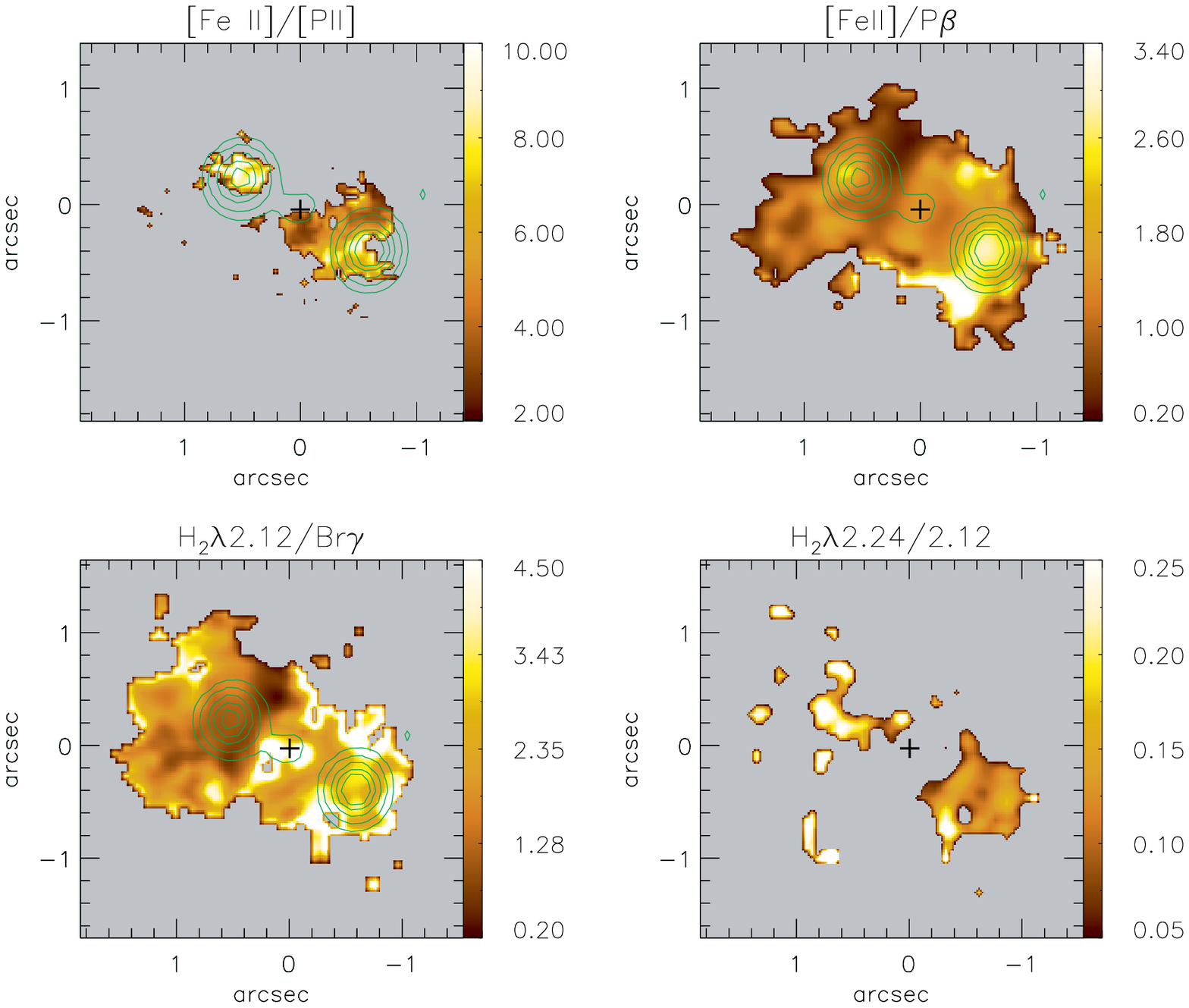}
\caption{Top: \feii$\lambda1.25\,\mu$m/\pii$\lambda1.8861\,\mu$m (left) and 
[Fe\,{\sc ii}]$\lambda$1.2570$\,\mu$m/Pa$\beta$ line ratio map (right). Bottom: 
H$_2\lambda$2.1218$\,\mu$m/Br$\gamma$  (left) and  \h2$\lambda2.2447\,\mu$m/\h2$\lambda2.1218\,\mu$m (right).
 The green contours overlaid to the \feii/\pii, \feii/\pb\ and \h2/\br\ maps are from the 6~cm radio image, the cross marks the position of the continuum peak.  The gray areas are masked regions following the same criteria used in the flux maps. The green contours overlaid to some panels are from the 6~cm radio image of \citet{ulvestad89} and the cross marks the position of the peak of the continuum emission. }
\label{ratio}
\end{figure*}


The excitation mechanisms of the \feii\ and \h2\ emission lines can be investigated using emission-line ratio maps. Figure~\ref{ratio} shows the 
\feii$\lambda1.2570\,\mu$m/\pii$\lambda1.8861\,\mu$m (top-left panel), \feii$\lambda1.2570\,\mu$m/\pb\ (top-right panel), \h2$\lambda2.1218\,\mu$m/\br\ (bottom-left panel) and \h2$\lambda2.2447\,\mu$m/\h2$\lambda2.1218\,\mu$m (bottom-right panel) line ratio maps. The first two are useful to investigate the \feii\ emission origin and the last ones to investigate the \h2\ emission origin. 

The \feii/\pii\ map shows values ranging from $\sim$2 to 10, with the highest values observed at the location of the northeastern radio knot and the lowest values at regions next to the nucleus of the galaxy. At the position of the southwestern radio knot,  \feii/\pii\ is $\sim$7. The \feii/\pb\ line ratio shows values ranging from 0.5 to up to 3.5, with the highest values observed in regions next to the southwestern radio knot. A small enhancement in this ratio is also observed next to the northeastern radio knot, where the values are  approximately 2.5. At most locations, the \feii/\pb\ ratio shows values in the range 0.6 to 2.0. Values smaller than 0.6 are observed only in a small region at 0\farcs5 north-northeast of the nucleus.

The \h2/\br\ line ratio shows values in the range from 0.2 to 4.5, with the highest values observed in regions surrounding the southwestern radio knot and at the nucleus. For most locations, the values of \h2/\br\ are in the range 0.6-2.0 and the smallest values of 0.3 are observed at 0\farcs5 north-northeast of the nucleus, at the same location where the \feii/\pb\ ratio map presents its smallest value. Finally, the \h2$\lambda2.24/2.12$ ratio map presents values ranging from 0.1 to 0.3 with the highest values seen at 0\farcs5 northeast of the nucleus, associated to the radio knot there.

\begin{figure}
\centering
\includegraphics[scale=0.7]{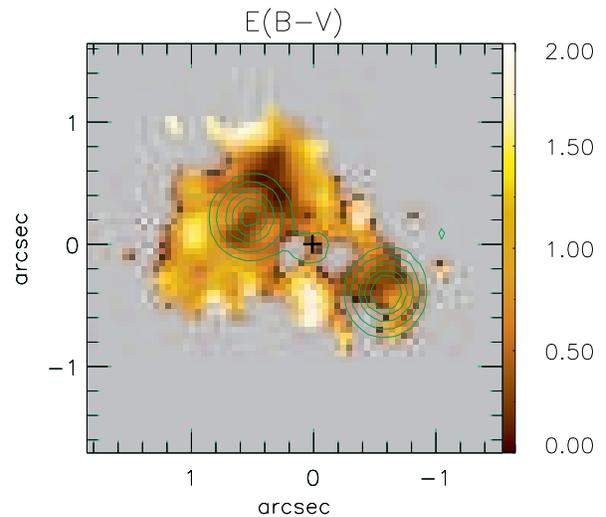}
\caption{Reddening map obtained from the Pa$\beta$/Br$\gamma$ line ratio. Green contours are from the radio image and the central cross marks the position of the nucleus.}
\label{ebv}
\end{figure}

In Figure~\ref{ebv} we present a reddening map obtained from the Pa$\beta$/Br$\gamma$ line ratio using
\begin{equation}
 E(B-V)=4.74\,{\rm log}\left(\frac{5.88}{F_{Pa\beta}/F_{Br\gamma}}\right),
\end{equation}
where $F_{Pa\beta}$ and $F_{Br\gamma}$ are the fluxes of $Pa\beta$ and $Br\gamma$ emission lines, respectively. We have adopted the intrinsic
 ratio $F_{Pa\beta}/F_{Br\gamma}=5.88$ corresponding to case B recombination \citep{osterbrock06} and used the reddening 
law of \citet{cardelli89}. The values of $E(B-V)$ for NGC\,5929 range from 0 to 2, with the highest values observed to the south, southeast and east of the nucleus, while smaller values are observed  approximately along the region covered bye the radio jet at $PA\sim45^\circ$, from  the northeast to the southwest of the nucleus.

\subsection{Stellar Kinematics}

The K-band spectra of Fig.~\ref{large} show clearly the CO absorption band heads around 2.3~$\mu$m. We used the Penalised
Pixel-Fitting ({\sc ppxf}) method of \citet{cappellari04} to fit these bands and obtain the line-of-sight velocity distributions (LOSVD) of the stars, using the Gemini library of late spectral type stars observed with the Gemini Near-Infrared Spectrograph (GNIRS) IFU and  NIFS \citep{winge09} as templates. The stellar LOSVD was approximated by a Gaussian distribution. The {\sc ppxf} outputs the  stellar radial velocity ($V_*$), the corresponding velocity dispersion ($\sigma_*$), as well as the uncertainties for both parameters at each spaxel. 

Figure~\ref{stel} presents the resulting maps for $V_*$ and $\sigma_*$. The signal-to-noise ratio of the spectra at locations next to the borders of the field of view was too low and we could not obtain good fits. These borders were then masked out. The white (black) regions in the $V_*$ ($\sigma_*$) map correspond to  these locations, where the uncertainties in  $V_*$ or $\sigma_*$  are larger than 30~\kms. The stellar velocity field shows a velocity amplitude of about 100~\kms, with redshifts to the northeast and blueshifts to the southwest, thus opposite from the rotation field observed for the gas, that has redshifts to the southwest and blueshifts to the northeast, as observed in Fig.~\ref{vel}. 
 The $\sigma_*$ map shows values ranging from 40 to 180~\kms\ with a  median value of $\sigma_*=114\pm7$~\kms\ and presents a partial ring of lower $\sigma_*$ values (60--100 \kms) with radius of 0\farcs5 surrounding the nucleus.

\begin{figure*}
\centering
\includegraphics[scale=0.9]{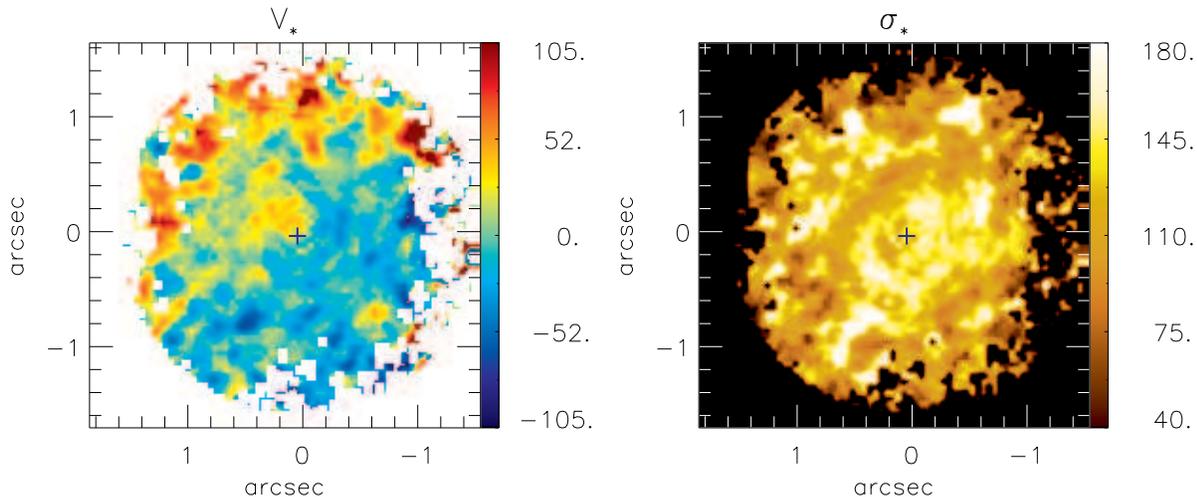}
\caption{Stellar velocity ($V_*$) field (left) and corresponding velocity dispersion ($\sigma_*$) map (right). The central cross marks the position of the nucleus, the color bars show $V_*$ and $\sigma_*$ values in units of \kms\ and the white/black regions in the $V_*$/$\sigma_*$ maps represent the locations where we could not get good fits of the galaxy spectra.}
\label{stel}
\end{figure*}

\subsection{Gas Kinematics}

We used the centroid wavelength of the emission lines [P\,{\sc ii}]\,$\lambda$1.1886\,$\mu$m, 
 [Fe\,{\sc ii}]\,$\lambda$1.2570\,$\mu$m, Pa$\beta$ and H$_2\,\lambda$2.1218$\mu$m at each position to map the velocities of the ionized gas forbidden lines, ionized gas permitted lines and molecular gas. These lines have been chosen because they present the highest S/N ratios among their species. 

These maps are shown in Figure~\ref{vel}, where the white regions have been masked due to bad fits, following the same criteria used for the flux distributions of Fig.~\ref{fluxmaps}.  All velocity maps show redshifts of up to 220\,\kms\ to the southwest of the nucleus and blueshifts of similar amplitude to the northeast of it. The zero velocity line is oriented approximately along $PA=-30^\circ/150^\circ$, coincident with the strip where the lines are double.

Figure~\ref{sig} presents the corresponding velocity dispersion ($\sigma$) maps. The $\sigma$ values for all emission lines are in the range 30 to 200\,\kms. At most locations, $\sigma$ is lower than 100\,\kms. The highest $\sigma$ values are observed along the SE-NW strip (that has a width of $\approx$ 0\farcs3, thus 50\,pc at the galaxy) crossing the nucleus perpendicularly to the radio jet and are due to the double components observed in the emission lines at these locations. These double components were attributed to and equatorial outflow in Paper I, where we presented the results only for \feii. Now we compare the \feii\ and \pii\ maps with those of \pb\ and \h2\ and we find some differences. The  \feii\ and \pii\  maps show that, besides the high $\sigma$ values observed along the SE-NW strip, there is also an increase of $\sigma$ at and around the locations of the radio hotspots. The \pb\ and \h2\  maps, on the other hand, do not show any evidence of increased $\sigma$ there.  A comparison between the profiles of these lines at the location of the radio hotspots can be seen in Fig.~\ref{profiles}: it is clear that the profile of the \feii\  line is broader than those of \pb\ and \h2\ lines as already reported in Paper I, being attributed to the interaction of the radio jet with the ambient gas.

\begin{figure*}
\centering
\includegraphics[scale=0.9]{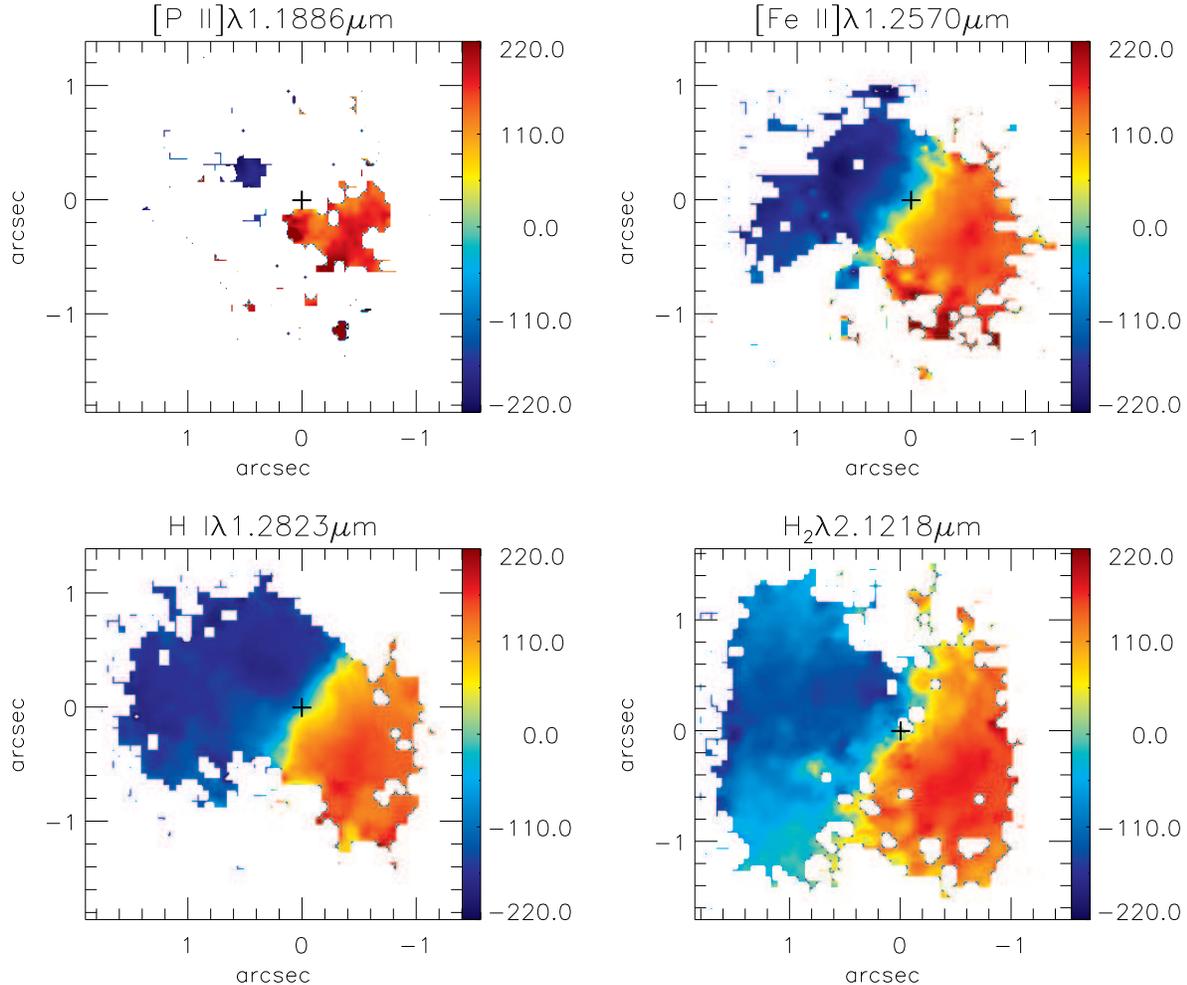}
\caption{Centroid velocity maps derived from th fit of the  [P\,{\sc ii}]\,$\lambda$1.1886\,$\mu$m, 
 [Fe\,{\sc ii}]\,$\lambda$1.2570\,$\mu$m, Pa$\beta$ and H$_2\,\lambda$2.1218$\mu$m emission-line profiles  using a single Gaussian.  The central cross marks the position of the nucleus and the scale of the color bar is shown in units of \kms.} 
\label{vel}
\end{figure*}

\begin{figure*}
\centering
\includegraphics[scale=0.9]{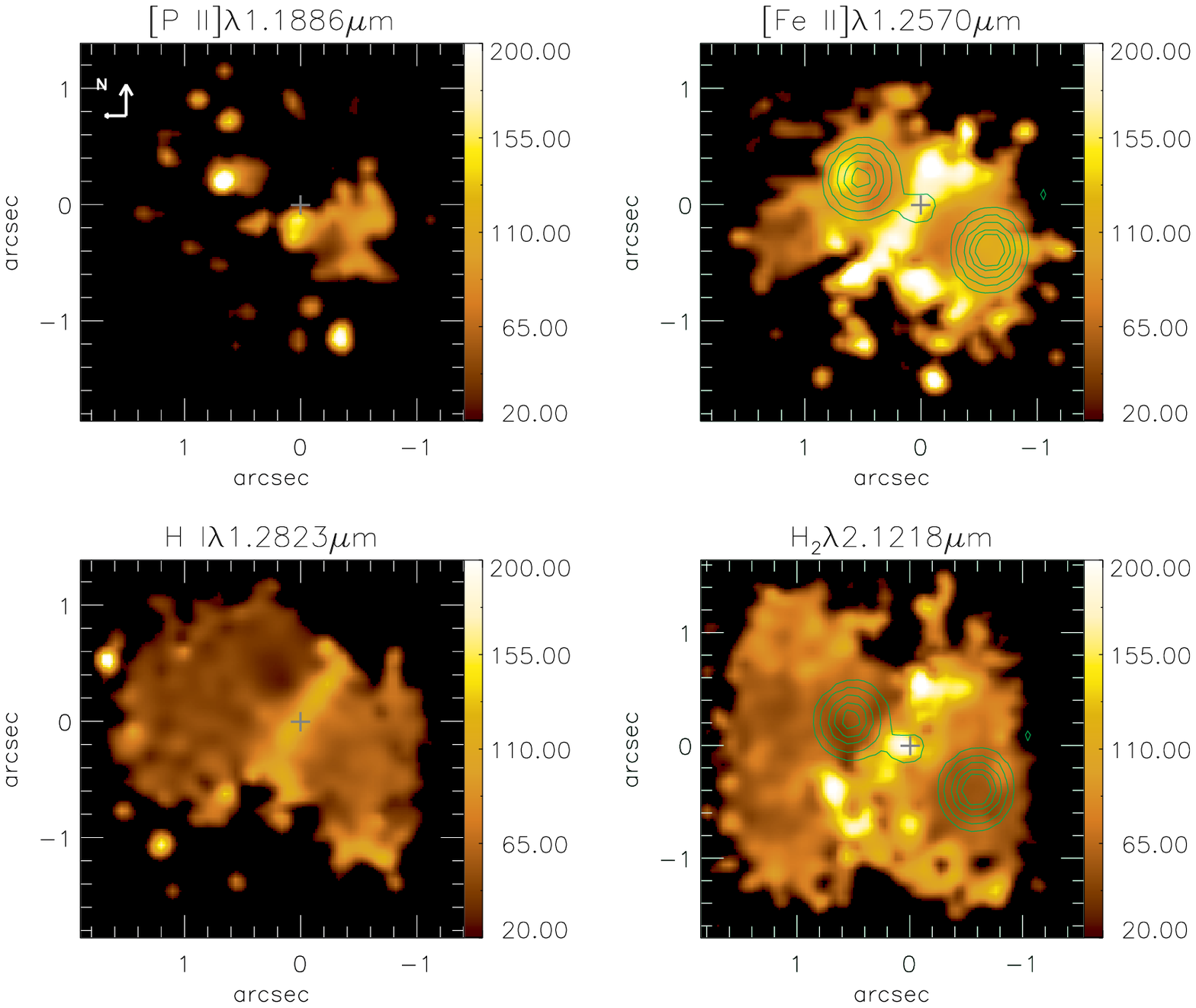}
\caption{Velocity dispersion maps derived from the [P\,{\sc ii}]\,$\lambda$1.1886\,$\mu$m, 
 [Fe\,{\sc ii}]\,$\lambda$1.2570\,$\mu$m, Pa$\beta$ and H$_2\,\lambda$2.1218$\mu$m emission line profiles.  The central cross marks the position of the nucleus, the scale of the color bar is in units of \kms\ and the green contours are from the 6\,cm radio image of \citet{ulvestad89}.}
\label{sig}
\end{figure*}

\subsection{Channel Maps}

In  figures  \ref{slicepb}, \ref{slicefe} and \ref{sliceh2} we show velocity-channel maps along the \pb, \feii\  and \h2\ emission-line profiles, respectively. 
The \h2\ channel maps show emission for a velocity range from $\sim-300$ to $300$~\kms, with the emission moving from northeast to southwest as the velocity increases, being consistent with a rotating disk. At low velocities, the \pb\ channel maps are similar to that observed for \h2, but additional \pb\ emission from high velocity gas is observed associated to the radio hotspots to the northeast at the blueshifted channels and to the southwest at at the redshift channels.  This suggests an interaction of the radio with the \pb\ emitting gas. For the \feii\ emission, the correlation of the line emission with the radio hotspots is more clear with emission being observed at velocities of up to $-$560~\kms\ at blueshifts and 500~\kms\ at redshifts. 

\begin{figure*}
\centering
\includegraphics[scale=0.8]{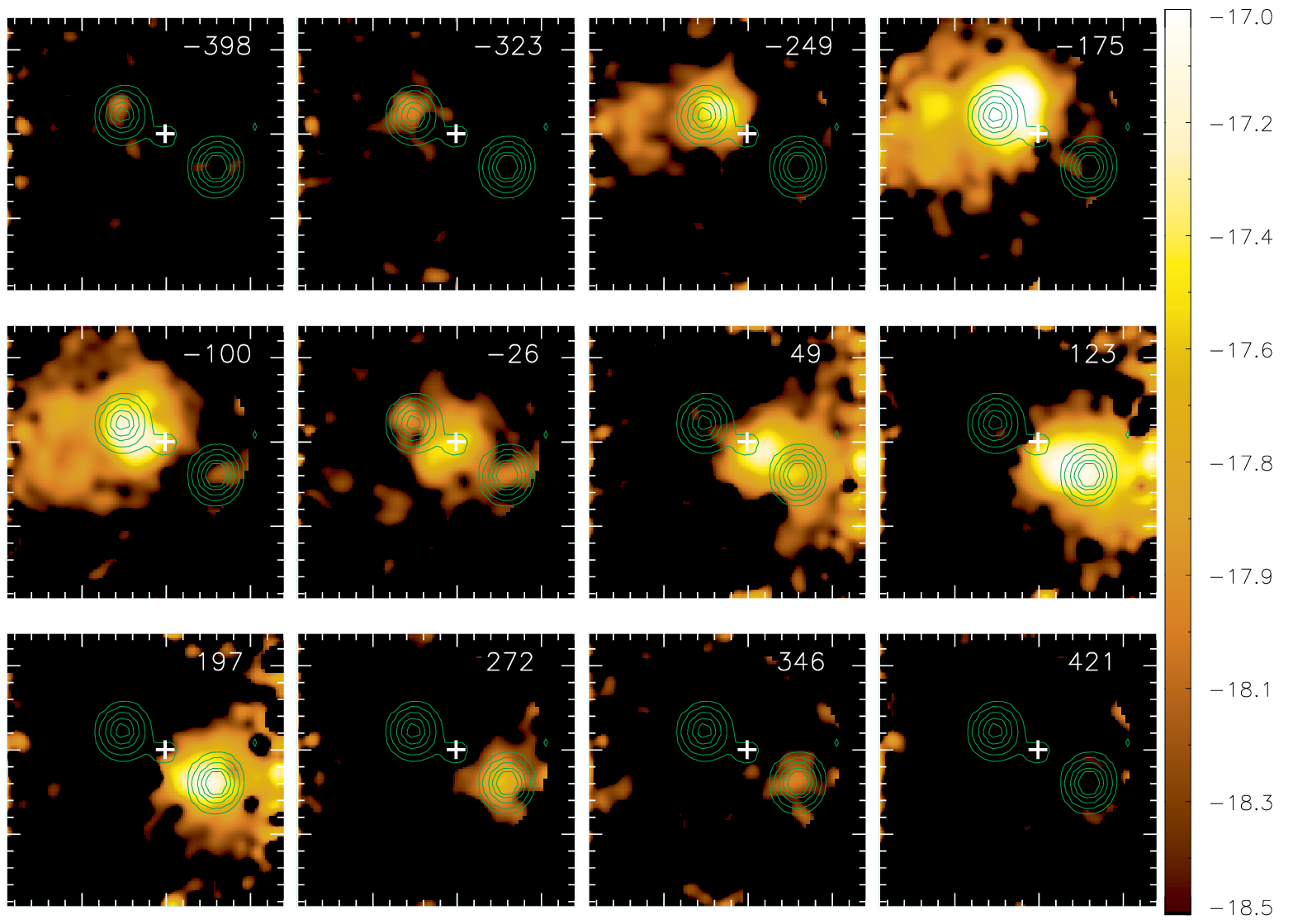}
\caption{Channel maps along the \pb\ line profile for a velocity bin of 75~\kms, corresponding to three spectral pixels. The corresponding velocity is shown in the top-right corner of each panel in units of \kms\ and the green contours are from the radio jet. The color bar shows the flux scale in logarithmic units.} 
\label{slicepb}
\end{figure*}

\begin{figure*}
\centering
\includegraphics[scale=0.8]{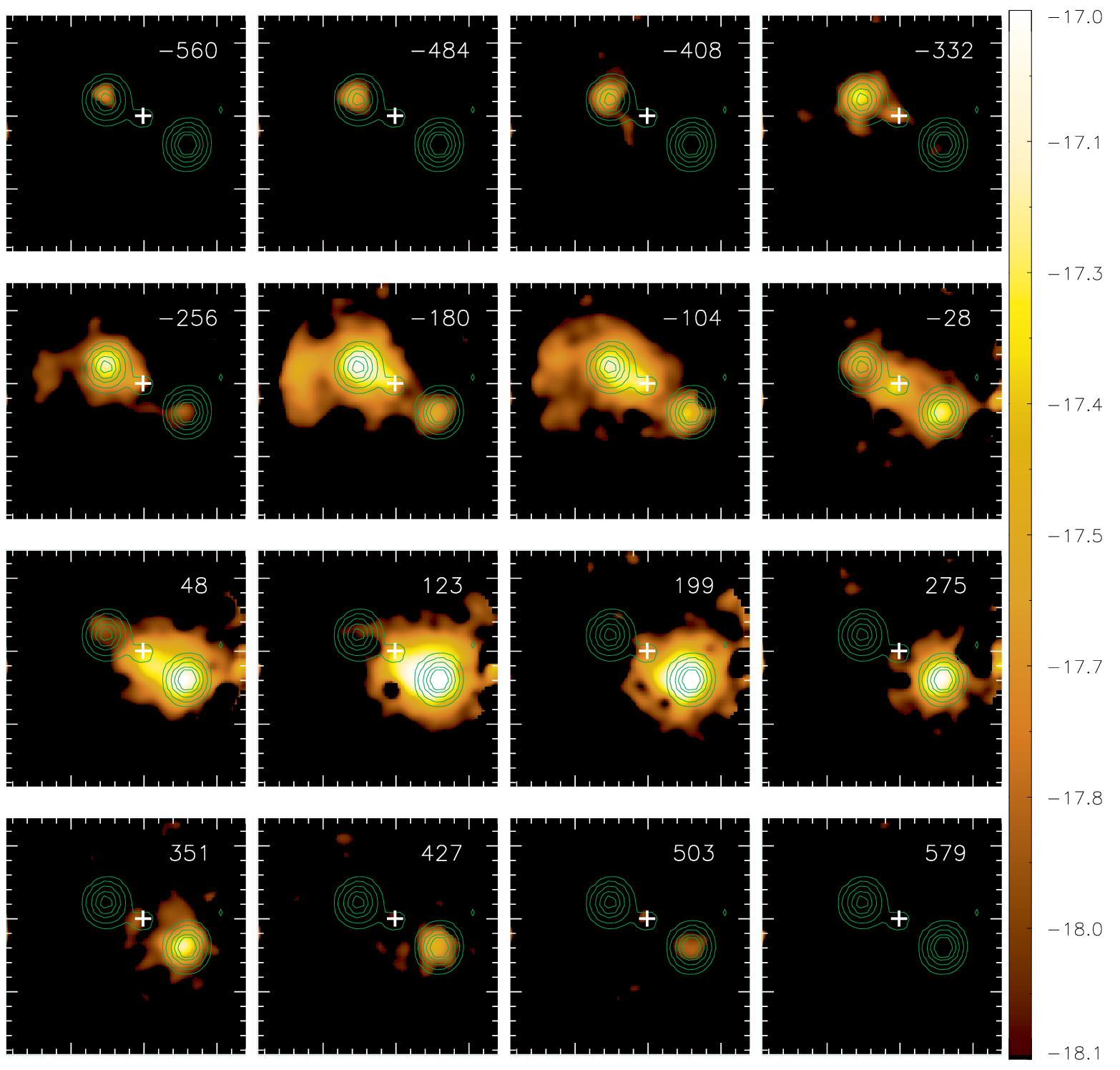}
\caption{Same as Fig.~\ref{slicepb} for the \feii$\lambda1.2570~\mu$m emission line.}
\label{slicefe}
\end{figure*}

\begin{figure*}
\centering
\includegraphics[scale=0.8]{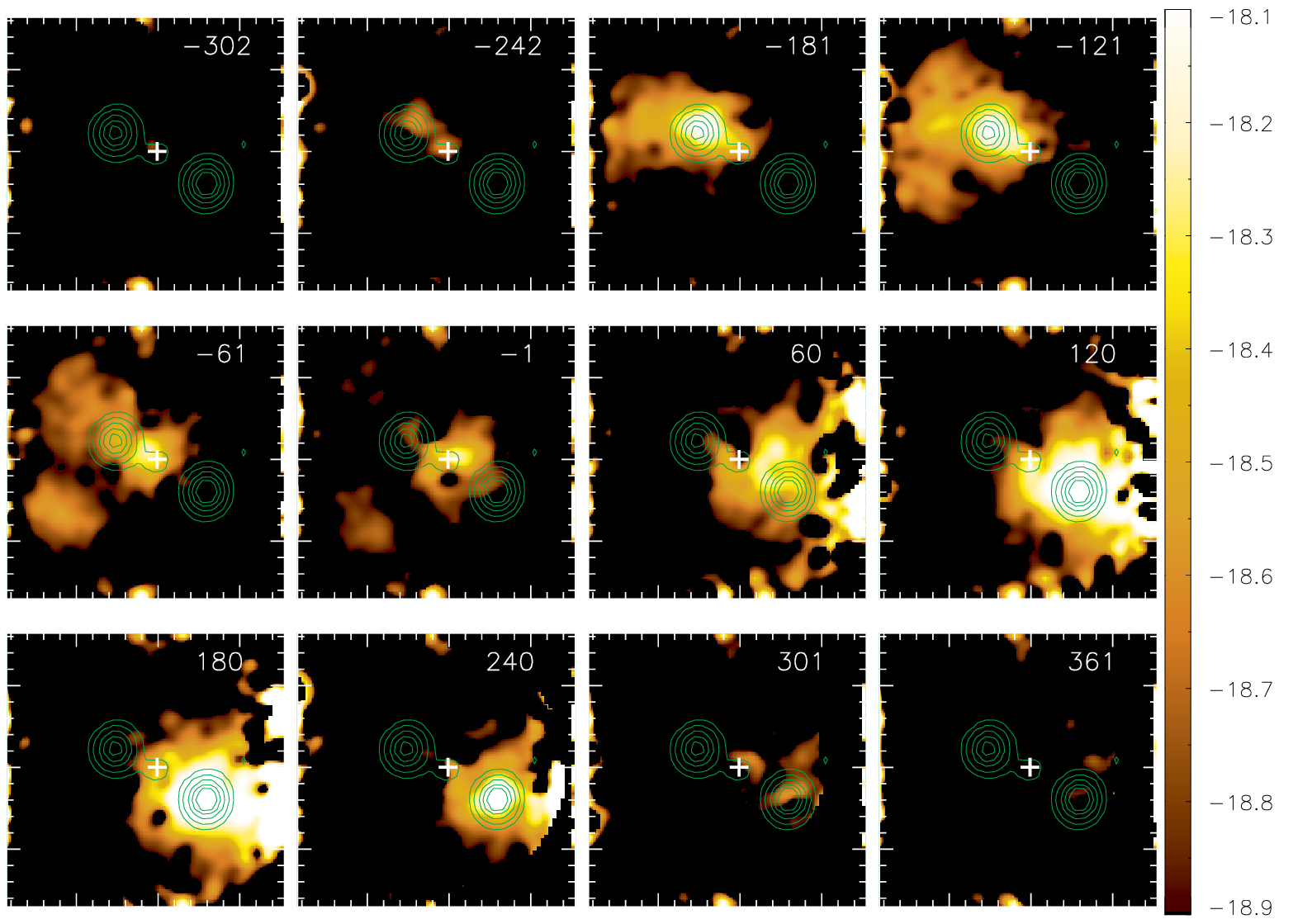}
\caption{Same as Fig.~\ref{slicepb} for the \h2$\lambda2.1218~\mu$m emission line for a velocity bin of 60 km/s, corresponding to two spectral pixels.}
\label{sliceh2}
\end{figure*}

\section{Discussion}
\subsection{The emission-line flux distributions}

All emission lines present their flux distributions more extended along the $PA=60/240^\circ$, which is the orientation of the radio jet \citep{ulvestad89,su96}. The companion galaxy NGC\,5930 is also at $PA=60$. Along the perpendicular direction, the line-emission extends only to $\sim$0\farcs7 from the nucleus for most lines (a bit more for H$_2$). In particular, along the SE-NW strip, we have concluded in Paper I that the gas is outflowing from the nucleus.

Although the bulk of the emission-line flux distributions is similar for all species, some differences are observed: (i) while the flux in the H\,{\sc i} recombination lines peak at 0\farcs5 northeast of the nucleus, the \feii, \pii\ and \h2\ lines peak at 0\farcs6 southwest of the nucleus; (ii) the \feii, \pii\ and H\,{\sc i} lines are more collimated along the radio axis than the H$_2$ emission. We have similar differences between the flux distribution (and kinematics) of the ionized and molecular gas for most of the Seyfert galaxies we have studied so far in this project  \citep{eso428,n4051,n7582,n1068-exc,mrk1066-exc,mrk1066-kin,mrk1157,sb09,sb10,schonel14}.

We also note that the flux values along the SE-NW strip are lower than along the direction of the radio jet, revealing that the outflowing gas discovered in Paper I does not contributes much to the total luminosity.  This is consistent with the interpretation that the radiation from the AGN escapes mainly along the ionization axis, which seems to be coincident with the radio axis and is blocked along the perpendicular direction due to the dusty torus postulated by the AGN unified model. This interpretation is supported by our discussion in Sec.~\ref{origin-double}, in which we conclude that X-ray heating (as X-rays can escape through the torus)  is the main excitation mechanism of the H$_2$ emission of both kinematic components observed along $PA=-30/150^\circ$.

We can compare our results with previous ones from the literature.
Optical Integral Field Spectroscopy (IFS) was obtained for NGC\,5929 by \citet{ferruit97} at a seeing of 0\farcs8 and by \citet{oasis} with the instrument OASIS at the Canada France Hawaii Telescope (CFHT) in the low spatial resolution mode (0\farcs4 sampling). \citet{ferruit97} present a contour map for the [N\,{\sc ii}]+H$\alpha$ emission, showing that it is more extended along the $PA$ of the radio jet and with the peak of emission observed at $\sim$0\farcs6 southwest of the nucleus, in good agreement with our flux distributions in the near-IR. \citet{oasis} present flux and kinematic maps for the \oiii$\lambda5007$, H\,$\beta$ and [N\,{\sc i}]$\lambda\lambda5198,5200$, finding a similar flux distribution to that of \citet{ferruit97}. Although the optical emission-line flux distributions are also more extended along  $PA=60/240^\circ$, similarly to what we observe, the optical lines do not show an emission peak to the northeast, as we have observed for the near-IR lines.  We attribute this differences to the worse spatial resolution of the optical observations. This interpretation is supported by the work of  \citet{rosario10}, who present a Hubble Space Telescope narrow-band image for the \oiii$+H\,\beta$ emission that clearly shows two main emission regions, one to the northeast and other to the southwest, with the highest fluxes observed to the southwest, in good agreement with the \feii, \pii\ and \h2\ flux distributions. 

Why do the near-IR H\,{\sc i} recombination lines present distinct flux distributions from that of \hb\ in the flux map of \citet{oasis}? As seen in Fig.~\ref{fluxmaps} the \pb\ and \br\ flux maps present their  peak at 0\farcs5 northeast of the nucleus with a secondary peak at 0\farcs6 southwest of it. On the other hand, the \hb\ flux distribution shows its emission peak to the southwest and although the emission is also extended to the northeast, there is no emission peak as observed for the near-IR lines. We attribute this difference to dust extinction, that is stronger in the optical than in the near-IR. Indeed, as seen in Fig.~\ref{ebv}, the reddening is somewhat larger to the east/northeast of the nucleus supporting the conclusion that the near-IR emission is probing a dustier region than the optical emission.  A difference of 0.5 mag in $E(B-V)$ between the northeast and southwest sides corresponds to a ratio of 2.2 between the Pa$\beta$ fluxes of the northeast and southwest using the \citet{cardelli89} extinction law, while for H$\beta$ the ratio would be higher, of $\approx$ 2.9. This difference would be enough to make the northeastern hotspot brighter than the southwestern hotspot in the \hb\ flux map of \citet{oasis}, supporting our conclusion that the difference between the near-IR and optical flux distributions is due to extinction.

\subsection{The stellar kinematics and the mass of the SMBH}

The stellar velocity field (left panel of Fig.~\ref{stel}) shows a rotation component, with the north-eastern side of the disk approaching and the south-western side receding with a velocity amplitude of about 80~\kms.  In order to derive the systemic velocity ($V_s$), orientation of the line of nodes ($\Psi_0$), inclination ($i$) and eccentricity ($e$) of the disk, we fitted the stellar velocity field using the {\sc diskfit} code \citep{spekkens07,sellwood10,kuzio12}.

The best rotating disk model is shown at the top-central panel of Fig.~\ref{diskmodel}. The observed velocity field is shown at the top-left panel and the residual map is presented at the top-right panel of the same figure, respectively.  The corresponding kinematic parameters are shown  in Table~\ref{kinpar}.

The stellar velocity dispersion ($\sigma_*$) can be used to estimate the mass of the supermassive black hole ($M_\bullet$) at the center of NGC\,5929, using the 
 $M_{\bullet}-\sigma$ relationship. We obtain $\sigma_*=133\pm8$~\kms\ using the pPXF for an integrated K spectrum over the whole field of view. The $M_\bullet$ can be estimated by \citep{msigma}:

\[
{\rm log}\left(\frac{M_\bullet}{10^9\,M_\odot}\right) = -(0.500\pm0.049)+(4.420\pm0.295){\rm log}\left(\frac{\sigma}{200\,\rm km s^{-1}}\right). 
\]
We estimated a mass of $M_\bullet=5.2^{1.6}_{-1.2}\times10^7~{\rm M_\odot}$.

The central velocity dispersion quoted in the Hyperleda database \citep{paturel03} is  $\sigma_*=120.6\pm12.9$~\kms, and results in $M_\bullet=3.4^{3.3}_{-1.9}\times10^7~{\rm M_\odot}$.

\subsection{The gas Kinematics}

\subsubsection{Equatorial outflow}

In Paper I we have reported the discovery of a peculiar gas outflow along the SE-NW strip that has a width of $\approx$\,50~pc and an extent of $\sim$300\,pc along $PA=-30/150^\circ$, perpendicularly to the radio jet. At locations away from the SE-NW strip, the gas kinematics is consistent with orbital motion in  a disk that is counter-rotating relative to the stellar disk. This rotation component is clear in our Fig.~\ref{vel} that is based on a single-Gaussian fit to the emission line profiles and can be compared to the stellar velocity field shown in the left-panel of Fig.~\ref{stel}.  Our stellar and gas velocity fields are very similar to those presented by \citet{oasis}, and derived from optical IFS, although they did not detect the outflow in the SW-NW strip due to their poorer spatial resolution.

In order to better show the distinct kinematic components along the SE-NW strip, where the mass outflow was observed in Paper I, we extracted one-dimensional cuts from the \feii\ flux distribution, velocity field and $\sigma$ map for the two components along the strip. The $\sigma$ of the two components was kept the same, as done in Paper I. These cuts are shown in Fig.\,\ref{onedfe}, where it can be seen that the velocity of the two components are almost constant along the strip, with values of $-150$ and 150\,\kms\  relative to the systemic velocity of the galaxy for the blue and red component, respectively. The fluxes of both components are similar at most locations, with the red component being brighter at distances smaller than 0\farcs5 from the nucleus to the southeast. 
 These components were attributed to the presence of an equatorial outflow (perpendicular to the radio jet) in Paper I. Such gas outflows appear in recent theoretical models of accretion disk winds \citep{li13} as well as in  outflowing torus models \citep[e.g][]{honig13,elitzur12}, in which the outflows are originated as a consequence of the conservation of the gas angular momentum. 

We can use the observed velocity and geometry of the emitting gas to estimate its mass outflow rate ($\dot{M}$) in the equatorial outflow. The double components observed along $PA=-30/150^\circ$ are attributed to the equatorial outflow from the nucleus. As the outflow is observed as two similar velocity components, one in redshift and the other in blueshift, we have considered a scenario in which the outflow has generated a hollow cylinder of outflowing mass. Considering the width of the SE-NW strip (50 pc) as being the height ($h$)  of the cylinder considered to have a radius $R_0$, the mass outflow rate can be estimated by

\begin{equation}
 \dot{M} = m_p N_e v f A,
\end{equation}
where $m_p$ is the proton mass, $N_e$ the electron density, $v$ is the velocity of the outflowing gas, $f$ is the filling factor and $A=2\pi R_0h$ is the lateral area of the cylinder. As the velocity of the outflow is approximately constant along the strip and there is no decreasing velocity as a function of distance from the center to NW and SW, we conclude that this cylinder has a large radius, at least the size of the region where emission is observed, $R_0>$0\farcs5. 

Assuming $N_e=500\,{\rm cm^{-3}}$, $f=0.01$  \citep[which are typical values estimated for other Seyfert galaxies -- e.g.][]{n2110-opt,n7213,m81,sb10},  $v=110$\,\kms\ (from Fig. \ref{onedfe}) and $R_0$=0\farcs5=87.5~pc, we obtain a lower limit (as the radius for the cylinder can be considered a lower limit)  to the mass outflow rate in ionized gas of $\dot{M} >  0.38~{\rm M_\odot\, yr^{-1}}$.  This value is in the range of mass outflow rates observed for other Seyfert galaxies \citep{crenshaw07,barbosa09,sb10,n7582,mrk1066-kin} with similar luminosities, although it is much smaller than those found in recent studies for high luminosity Seyfert 2 galaxies  (${\rm log(L[O III]5007/L_\odot)} > 8.7$) which show $\dot{M}_{\rm out} \approx 370 - 2700~{\rm M_\odot\, yr^{-1}}$ \citep[e.g.][]{mcelroy14}. 

Finally, due to the apparent unique  velocity of the outflow, it looks like that it is not continuous, but was generated by a "blast" that has produced an expanding (cylindrical) shell of gas. If this is the case, we can estimate the age of the ``blast" by  $\tau=R_0/v$ resulting in $\tau=0.8$~Myr. Actually, this can be considered an upper limit for the age, as the outflow may have decelerated since the ejection from the AGN.

\begin{figure}
\centering
\includegraphics[scale=0.55]{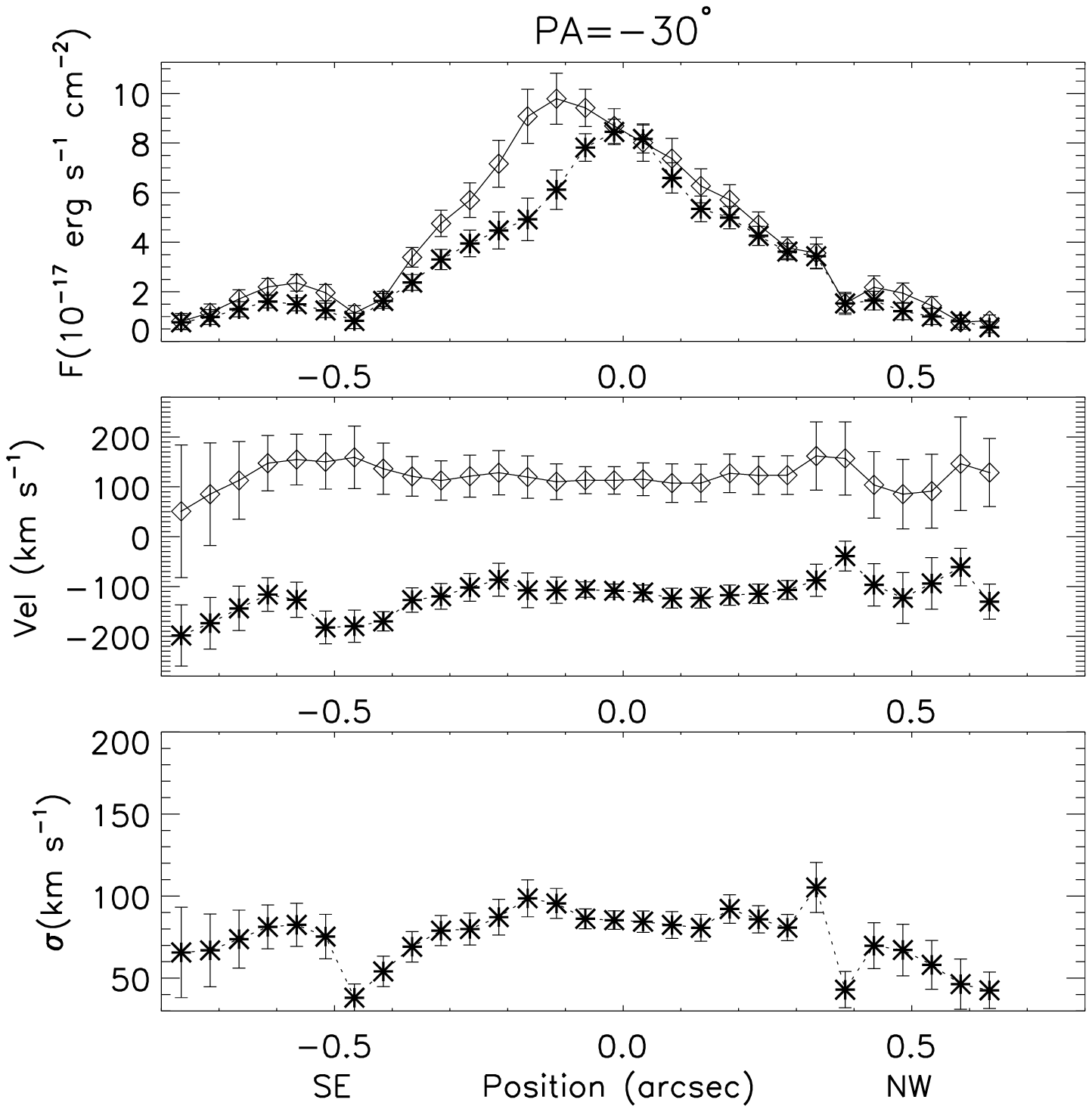}
\caption{One-dimensional cut for the fluxes (top), velocity (middle) and velocity dispersion (bottom) from two-Gaussian components fit to the \feii\ emission-emission line profile along the SE-NW strip. The velocity dispersion was contained to be the same for both components during the fits.}
\label{onedfe}
\end{figure}

\subsubsection{Interaction between the radio jet and emitting gas}

Our data allowed us to observe also the kinematic effects of the interaction of the radio jet with the \feii\ emitting gas, as discussed in Paper I based on a two Gaussian component fit to the \feii\ line profile. The interaction of the radio jet with the gas produces wings in the \feii\ line profile at locations around the radio hotspots. In the single-Gaussian fit, this effect appears as an enhancement of $\sigma$ at these locations, as seen in Fig.~\ref{sig}.  \citet{rosario10} claims also to have found similar signatures of shocks due to the radio jet, using optical long slit observations with the HST. 

Although  signatures of the interaction of the radio jet with the ISM are clearly observed for the \feii\ emission, the \h2\ and H recombination lines show weaker signatures, only observed in the channel maps.  This indicates that the \feii\ emission traces a more disturbed kinematics than the molecular and atomic hydrogen. This can be interpreted as due to shocks destroying dust grains in its passage through the gas, releasing the Fe atoms, that are then ionized and emit.This conclusion is supported by the reduced reddening observed along the ionization and radio jet axis of the AGN, and also in agreement with previous results we have found for other Seyfert galaxies: the molecular gas and the ionized gas emitting the \feii\ lines trace distinct kinematics and flux distributions with the former more restricted to the plane of the galaxies and the latter usually associated to outflows from the nucleus \citep{n4051,n7582,mrk1066-exc,mrk1066-kin,mrk79,n1068-exc,sb09,sb10,n1068-kin}. 

We can also speculate on the age of the radio jet. Assuming the jet velocity of $0.1\,c$, if the jet were in the plane of the sky with a size of 0\farcs5, the age would be 2850 yr. If the inclination of the jet relative to the plane of the sky is 45$^\circ$, the age would be about 4000 years and we thus conclude that the radio jet is younger than the "blast" generating the equatorial outflow. Alternatively, the jet is much more inclined toward us and could have been emitted together with the "blast", taking also in consideration that the velocity of the outflow may have been larger when it was generated.






\subsubsection{Rotating disk}

As observed in the stellar velocity field, the  gas velocity fields (Fig.~\ref{vel}) also support the presence of a rotating disk component, although the gas is clearly counter-rotating relative to the stars. This was already observed by \citet{oasis}, as mentioned above. In order to confirm the presence of this component and derive the corresponding kinematic parameters, we fitted the \pb, \feii\ and \h2\  velocity fields using the {\sc diskfit} code \citep{spekkens07,sellwood10,kuzio12} for a rotating disk model. In order to fit the rotation component of the gas velocity field, we excluded the velocities from the SE-NW strip.  The exclusion followed the same criteria adopted in Paper I, in which regions with $\sigma > 100$~\kms\ for the \feii\ line along $PA=-30/150^\circ$ were attributed to the equatorial outflow. 

The corresponding best models are shown in the central panels of Fig.~\ref{diskmodel}, while the observed velocity fields and the residual maps are presented in the left and right panels, respectively. As the kinematics for distinct emission lines are similar, we show only the maps for the \h2\ in this figure. It can be seen that the residuals are smaller than 50~km/s at most locations, indicating that the velocity fields are reasonably well representing by rotation in a disk.  In Table~\ref{kinpar} we present the kinematic parameters for the best fit models. The systemic velocity ($V_s$), ellipticity of the orbits ($e$), inclination of the disk ($i$) and position of the kinematical center ($X_{cen}$, $Y_{cen}$) for the molecular and ionized gas and stars are very similar. The kinematical center is measured  relative to the peak of the K-band continuum and was kept fixed for the stellar velocity field in order to reduce the number of free parameters to be fitted, as the stellar velocity field is nosier than those for the emission lines. 

The orientation of the line of nodes ($\Psi_0$) obtained from the fit of the gas velocity fields are consistent with each other, while $\Psi_0$ for the stellar velocity field is approximately opposite to that of the gas (differing by 190$^\circ$).

We can compare the kinematical parameters from Table~\ref{kinpar} with those for the large scale disk, that are quoted in the Hyperleda database \citep{paturel03}. The heliocentric systemic velocity obtained from the fit is in good agreement with the large scale value ($V_s\approx2503$~\kms) obtained from optical emission lines, while the inclination of the disk and the ellipticity of the orbits are a bit smaller than those for the large scale disk ($i\approx24^\circ, ~ e=(1-b/a)\approx0.09$). The orientation of the line of nodes is $10-20^\circ$ smaller than that of the large scale disk \citep[e.g.][]{schmitt97}.  We attribute the origin of this rotating gas disk to the interaction with NGC5930.  The gas velocity amplitude, corrected by the inclination of the disk is about 580\,\kms.The high rotation velocities indicate that this gas is still not in orbital equilibrium in the galaxy gravitational potential. This gas is probably the source of the feeding of the AGN.

\begin{figure*}
\centering
\includegraphics[scale=0.7]{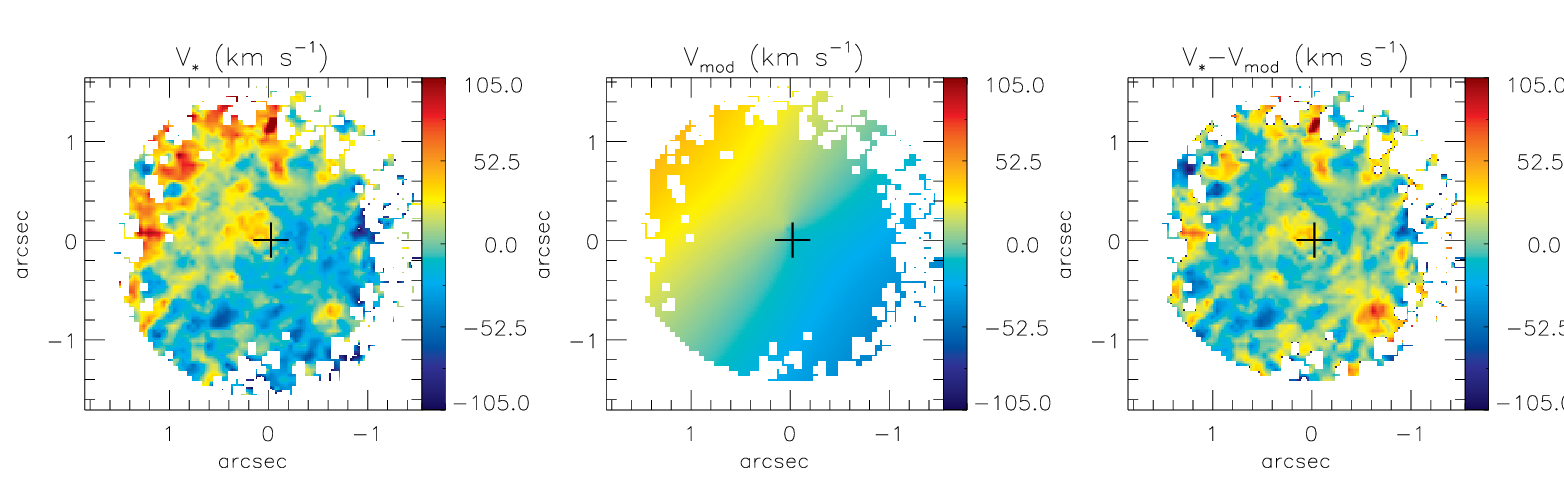}
\includegraphics[scale=0.7]{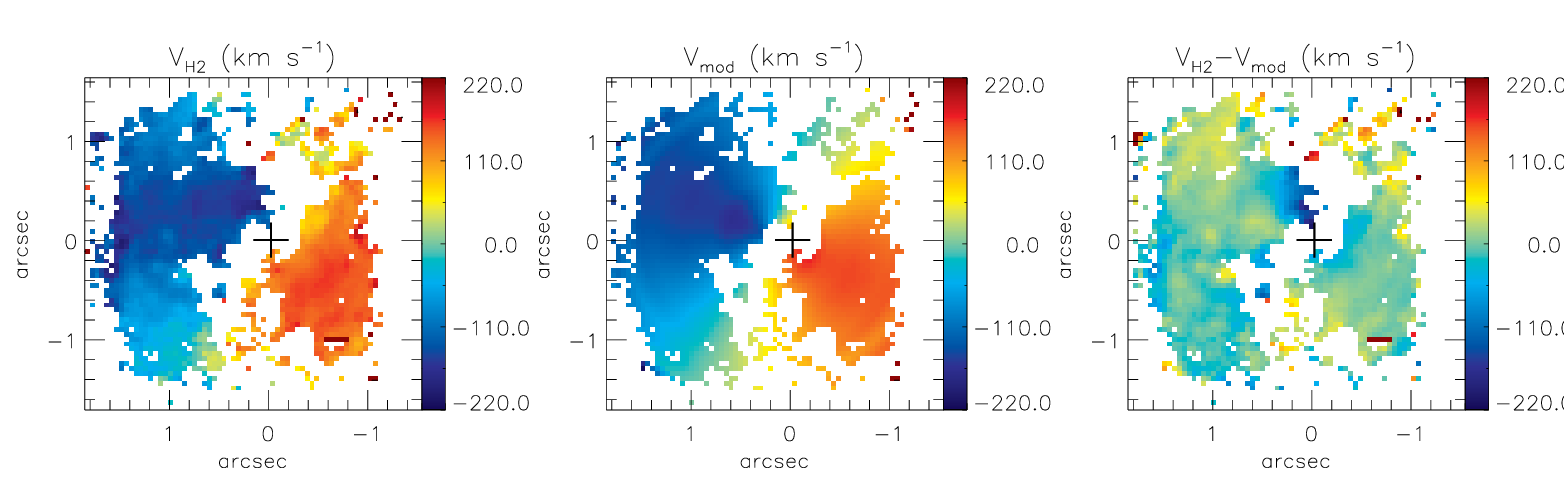}
\caption{From left to right: observed velocity field, rotating disk model and residual map (observed -- model) for the stellar (top) and \h2\ (bottom) velocity field. The color bars show the range of velocities in units of \kms\ relative to the systemic velocity of the galaxy and the central cross marks the position of the nucleus.}
\label{diskmodel}
\end{figure*}

\begin{table*}
\centering
\caption{Kinematic parameters from the best rotating disk model. Col 1: Fitted velocity field, col 2: heliocentric systemic velocity, col 3: orientation of the line of nodes, col 4: ellipticity, col 5: inclination of the disk and cols 6 and 7: position of the kinematical center relative to the peak of the K-band continuum emission.}
\vspace{0.3cm}
\begin{tabular}{l c c c c l }
\hline
Velocity Field & $V_s$ (\kms) & $\Psi_0$ & $e$ & $i$ & ($X_{cen}$, $Y_{cen}$) \\
\hline
H$_2$          & 2500.1$\pm$17.4  & 25.7$^\circ\pm$1.3$^\circ$ & $0.05\pm0.02$ & 18.4$^\circ\pm$1.2$^\circ$ & (0\farcs2$\pm$0\farcs1, $-$0\farcs1$\pm$0\farcs1) \\ 
Pa$\beta$      & 2472.8$\pm$18.5  & 36.5$^\circ\pm$1.1$^\circ$ & $0.05\pm0.02$ & 18.2$^\circ\pm$1.3$^\circ$ & ($-$0\farcs1$\pm$0\farcs1, 0\farcs0$\pm$0\farcs1) \\
\feii\         & 2469.3$\pm$18.5  & 25.4$^\circ\pm$4.4$^\circ$ & $0.05\pm0.04$ & 18.2$^\circ\pm$7.8$^\circ$ &(0\farcs1$\pm$0\farcs1, 0\farcs0$\pm$0\farcs1) \\
Stars          & 2489.4$\pm$17.1  & 219.9$^\circ\pm$1.3$^\circ$ & $0.05\pm0.03$ & 18.2$^\circ\pm$6.7$^\circ$ & (0, 0) fixed \\
\hline
\end{tabular}
\label{kinpar}
\end{table*}

\subsection{Mass of ionized and molecular gas}

The mass of the ionized gas in the inner 520$\times$520~pc$^2$ can be estimated by 
\begin{equation}
 M_{HII}\approx3\times10^{17}\left(\frac{F_{\rm Br\gamma}}{\rm erg\,s^{-1}cm^{-2}}\right)\left(\frac{d}{\rm Mpc}\right)^2 ~~~~~~[{\rm M_\odot}],
\end{equation}
where  $F_{\rm Br\gamma}$ is the integrated flux for the \br~emission line and $d$ is the distance to NGC\,5929 \citep{osterbrock06,sb09}. 
We have assumed an electron temperature $T=10^4$\,K and electron density $N_e=10^2\,{\rm cm^{-3}}$.  

 The mass of the warm molecular gas is given by
\begin{equation}
 M_{H_2}\approx5.0776\times10^{13}\left(\frac{F_{H_{2}\lambda2.1218}}{\rm erg\,s^{-1}\,cm^{-2}}\right)\left(\frac{d}{\rm Mpc}\right)^2~~~~~~[{\rm M_\odot}],
\label{mh2}
\end{equation}
where  $F_{H_{2}\lambda2.1218}$ is the integrated flux for the \h2$\,\lambda2.1218\,\mu$m emission line
and we have used the vibrational temperature T=2000\,K \citep[e.g.][]{scoville82,n4051}.

Integrating over the whole IFU field  (520$\times$520~pc$^2$) we obtain $F_{\rm Br\gamma}=3.4\pm0.4\times10^{-15}\,{\rm erg\,s^{-1}\,cm^{-2}}$ and
 $F_{H_{2}\lambda2.1218}\approx7.3\pm0.9\times10^{-15}\,{\rm erg\,s^{-1}\,cm^{-2}}$,  that results in  $M_{HII}=1.3\pm0.2\times10^6\,{\rm M_\odot}$
  and $M_{H_2}=471\pm58~{\rm M_\odot}$.

The masses of warm molecular and ionized gas are in the range of values found for other active galaxies, which have $M_{HII}=10^4-10^7~{\rm M_\odot}$ and $M_{H_2}=10^1-10^3~{\rm M_\odot}$  \citep{n4051,n7582,mrk1066-exc,mrk1157,n1068-exc,sb09,mrk79,mazzalay13}. However, the galaxies from these studies are located at a large range of distances (10--100 Mpc) and thus the corresponding sizes of the regions covered by the observations -- obtained with NIFS (and SINFONI) --  are distinct for the distinct objects. 

In order to be able to compare similar quantities, we now calculate the average mass surface density, instead of the mass. For NGC\,5929 we get $\sum_{M_{H_2}}=1.7\times10^{-3}{\rm M_\odot/pc^2}$ and $\sum_{M_{HII}}=4.8~{\rm M_\odot/pc^2}$ for the mass surface density of the warm molecular and ionized gas, respectively.

 These values are in the range observed for other galaxies, that renge from $\sum_{M_{H_2}}=2.4\times10^{-4}{\rm M_\odot/pc^2}$ for NGC~1068 \citep{n1068-exc} to  $\sum_{M_{H_2}}=7.1\times10^{-3}~{\rm M_\odot/pc^2}$ for NGC\,2110 \citep{diniz} for the warm molecular gas and from $\sum_{M_{HII}}=1.8\times10^{-1}{\rm M_\odot/pc^2}$ for NGC~1068 \citep{n1068-exc} to  $\sum_{M_{HII}}=42.8~{\rm M_\odot/pc^2}$ for NGC\,4151 \citep{sb09} for the ionized gas.  However, as the galaxies of our sample present distances in the range of 9.3 Mpc (for NGC\,4151) to 93.8 Mpc (for Mrk\,79), distinct average 
densities are expected for similar radial density profiles with decreasing gas densities as a function of distance from the nucleus.

In order to compare the masses of ionized and and warm molecular gas closest to the nucleus, we used the previous data obtained by our group to estimate the masses within a fixed physical aperture of $100\times100$\,pc$^2$. This aperture was chosen as being the whole field of view of the nearest object of the sample. The mass of ionized gas ionized gas within this aperture is in the range  $0.3-27\times10^5$\,M$_\odot$ and the mass of warm molecular ranges from 3 to 455~M$_\odot$, approximately. The lowest values for both ionized and molecular gas masses are observed for NGC\,4051, while NGC\,1068 presents the highest values. For NGC\,5929 we find $M_{HII}=3\pm0.5\times10^5\,{\rm M_\odot}$  and $M_{H_2}=47\pm6~{\rm M_\odot}$ for the inner $100\times100$\,pc$^2$, which corresponds to an aperture of only 0\farcs6$\times$0\farcs6.  The median ratio between ionized and molecular gas masses is $6.9\times10^3$, similar to the ratio for NGC\,5929 of $\sim6.4\times10^3$.

However, the total mass of molecular gas, including the cold gas is much larger than the values obtained here. The ratio between cold and warm molecular gas observed in the central region of active galaxies is in the range 10$^{5}-$10$^{7}$ \citep{dale05,ms06,mazzalay13} and thus, the total amount of molecular gas in the inner $3^{\prime\prime}\times3^{\prime\prime}$ of NGC~5929 should be at least $4.7\times10^7~{\rm M_\odot}$ with a surface mass density of  $170~{\rm M_\odot/pc^2}$.

\subsection{The origin of the \feii\ emission}

The origin of the \feii\ emission in AGNs can be investigated using the line-ratio maps shown in Fig.~\ref{ratio}. 
The \feii\ emission is excited in partially ionized gas regions, which can be produced by 
 X-ray \citep[e.g.][]{simpson96} and/or shock heating \citep[e.g.][]{forbes93}  of the gas. Values of 
 \feii/\pb$>2.0$ indicate that most of the \feii\ line emission is produced by shocks, while  
\feii/\pb$<0.6$ indicate that photo-ionization dominates the emission 
 \citep{rodriguez-ardila04,rodriguez-ardila05a}.

The \feii$\,\lambda1.2570\mu$m/\pb~line ratio  for NGC\,5929 is shown in the top-right panel of Fig.~\ref{ratio}. 
At most positions this ratio has values larger than 0.6, with the exception of only a small region at 
$\sim$0\farcs8 north-northeast of the nucleus.  At locations co-spatial with the radio knots, the  \feii$\,\lambda1.2570\mu$m/\pb ratio
 is larger than 2.0, suggesting that the interaction of the radio jet with the gas has an important
 role in the observed  \feii\ emission. 
At the northeastern radio hotspot,  \feii/\pb$\approx2.3$, while at the southwestern
 radio hotspot  \feii/\pb$\approx2.8$. At the nucleus and along THE SE-NW strip a typical value for  \feii/\pb\ is 
1.5, suggesting that both, photo-ionization and shocks contribute to the excitation of \feii, with a larger contribution of 
shocks though (as this value is closer to 2.0 than to 0.6).

The [Fe\,{\sc ii}]$\lambda$1.2570$\,\mu$m and [P\,{\sc ii}]$\lambda$1.8861$\,\mu$m lines  have 
similar excitation  temperatures, and their parent ions
have similar ionization potentials and radiative recombination coefficients and the \feii/\pii\ line ratio map 
shown in top-left panel of Fig.~\ref{ratio} is also useful to investigate the \feii\ emission origin. Values  
larger than 2.0  indicate that shocks have passed
through the gas destroying the dust grains, releasing the
Fe and thus enhancing its observed abundance. For supernova remnants, 
where shocks are the dominant excitation mechanism, [Fe\,{\sc ii}]/[P\,{\sc ii}]
 is typically  higher than 20 \citep{oliva01}. In NGC\,5929, at all locations where both lines are detected, 
\feii/\pii\ is larger than 2.0, suggesting that shocks contribute to the \feii\ emission and supporting the conclusion reached from the 
\feii/\pb\ ratio map. At the position of the hotspots, values of up to 10 are observed, supporting an even stronger interaction of the radio jet and with the gas.

As the \feii\  emission-line profiles are complex at many locations, as seen in Fig.~\ref{profiles}, with extended wings at locations close 
to the radio knots and double components along the SE-NW strip, we also constructed  \feii/\pb\ line-ratio channel maps (constructing line ratios at different velocity bins)
in order to determine the gas excitation mechanism at distinct velocities. The resulting channel maps
 are shown in  Fig.~\ref{chamapFePb}. Each panel presents the  \feii/\pb\ line-ratio map centered at the velocity shown in the top-right corner of the panel, 
relative to the systemic velocity of the galaxy. In order to avoid spurious features, we masked regions in which one or both lines present fluxes smaller than 3 times
 the standard deviation of the continuum next to the line. These regions are shown in gray in the figure and the contours are from the radio image. 

The \feii/\pb\  ratio shows 
values ranging from 0.2 to up to 4.5. The highest values are observed for the gas at the highest velocities  (largest blueshifts and redshifts) and present an excellent correlation with the
radio knots. These high \feii/\pb\  values are interpreted as a strong evidence that the \feii\ is produced by the interaction of the radio jet with the NLR gas. For smaller velocities, at most locations the  values are smaller than 2.0, with typical values of $\sim$1.2, suggesting that X-rays from the central AGN are the main drivers of the \feii\ emission. Line emission is observed along the SE-NW strip only at velocities lower than 180\,\kms. The values of \feii/\pb\ there are somewhat higher the one observed at other locations, but still much smaller than those seen at the radio knots. Typical values are \feii/\pb$=2-3$, suggesting that shocks are also important along the SE-NW strip. These shocks may be associated to the equatorial outflows we have reported at this locations in Paper I.

We conclude that the main excitation mechanism of the \feii\  emission in NGC\,5929 are shocks due to the radio jet, with some contribution from X-ray heating at locations distant from the radio structures. The highest velocity gas is associated to the interaction of the radio jet with the ISM and thus we conclude that its \feii\ emission is produced mainly by shocks, while X-ray excitation is more important for the rotating gas at lower velocities. These results are supported by the line-ratio maps and by the correlation between the emission-line flux distributions and the radio image, and are in good agreement with those of previous similar studies for other Seyfert galaxies \citep[e.g.][]{rodriguez-ardila04,
rodriguez-ardila05a,eso428,mrk1066-exc,sb99,sb09}.

\begin{figure*}
\centering
\includegraphics[scale=0.74]{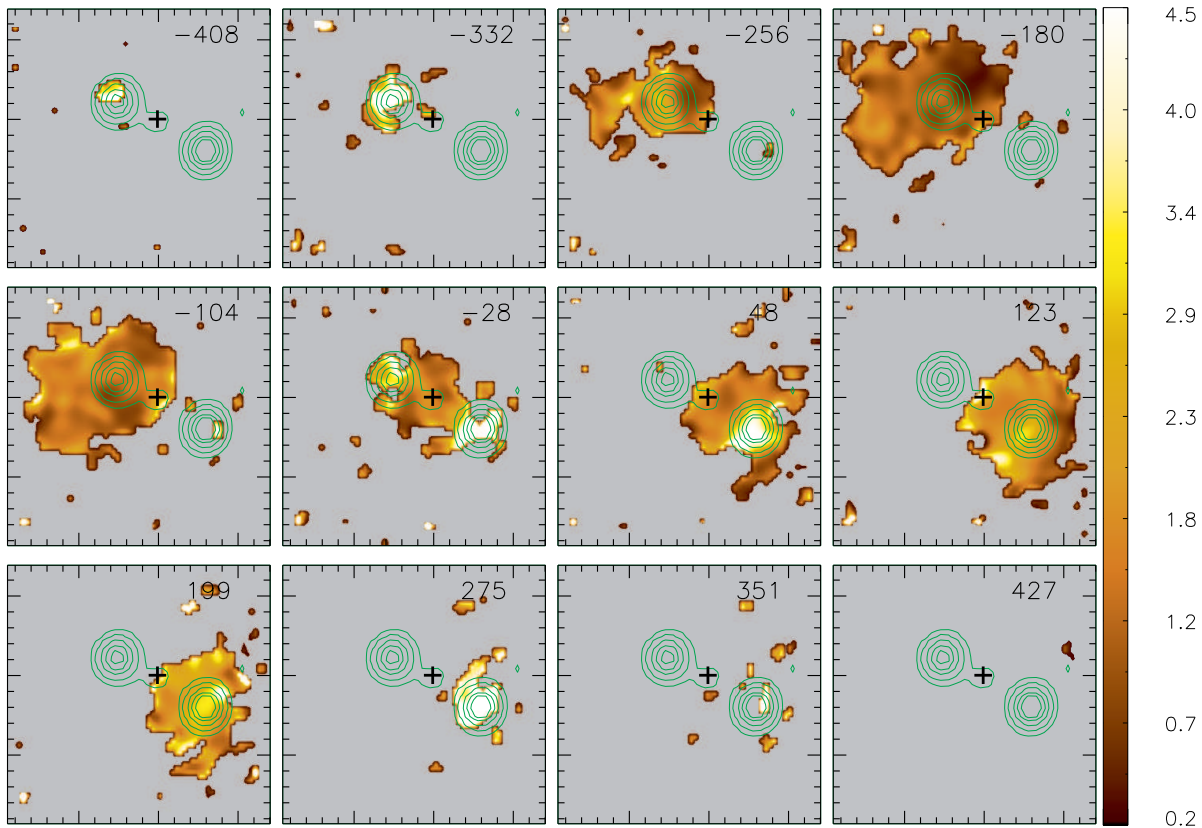}
\caption{Velocity channel maps for the [Fe\,{\sc ii}]$\lambda$1.2570$\,\mu$m/\pb\ line ratio. Each panel shows the fluxes centered at the velocity shown at its top-right corner relative to the systemic velocity of the galaxy. Grey regions are locations where one or both lines present fluxes smaller that 3$\sigma$ (the standard deviation of the continuum next to the line) and the green contours are from the radio image of \citet{ulvestad89}.}
\label{chamapFePb}
\end{figure*}

\subsection{H$_2$ excitation} \label{disc_h2}

The excitation of the H$_2$ line emission in the near-IR can be due to (i) fluorescence through absorption of soft-UV photons (912--1108 \AA) in
the Lyman and Werner bands \citep{black87} and/or by thermal processes due to the heating of the gas by shocks (due to the interaction of the radio jet with the emitting gas and/or supernovae) \citep{hollenbach89} or by X-rays from the central AGN  \citep{maloney96}. Several works have been aimed to study the origin of the H$_2$ emission in AGN using long-slit \citep[e.g.][]{reunanen02,rodriguez-ardila04,rodriguez-ardila05a,davies05,ramos-almeida09} and integral-field spectroscopy \citep[e.g.][]{eso428,n4051,n1068-exc,mrk1066-exc,mrk79,mrk1157,sb09,ms09, hicks09,friedrich10,mazzalay13,iserlohe13}.

In order to distinguish between fluorescence and thermal processes, we used the observed fluxes for all \h2\ emission lines in the K band to calculate the thermal excitation temperature.
If the H$_2$ emission is dominated by thermal processes and  under the assumption of an
{\it ortho:para} abundance ratio of 3:1, the following relation is valid \citep{wilman05}:

\begin{equation}
 {\rm log}\left(\frac{F_i \lambda_i}{A_i g_i}\right)={\rm constant}-\frac{T_i}{T_{\rm exc}},
\end{equation}
where $F_i$ is the flux of the $i^{th}$ H$_2$ line, $\lambda_i$ is its wavelength, 
$A_i$ is the spontaneous emission coefficient, $g_i$ is the statistical                         
weight of the upper level of the transition, $T_i$ is the
energy of the level expressed as a temperature and $T_{\rm exc}$ is the excitation temperature. Thus, if the observed fluxes can be reproduced by the equation above, the H$_2$ may be dominated by thermal processes. The resulting plot for $N_{\rm upp}=\frac{F_i \lambda_i}{A_i g_i}$ (plus an arbitrary constant) $vs$ $E_{\rm upp}={T_i}$  is shown in Fig.~\ref{excitation} for the fluxes shown in Table~\ref{fluxes} for the locations of the two radio hotspots at 0\farcs5 northeast and 0\farcs6 southwest of the nucleus. We do not show plots for other positions of Table~\ref{fluxes} because only few $H_2$ lines were detected at these locations.  As seen in Fig.~\ref{excitation}, the observed fluxes are well reproduced by the equation (shown as a continuum line) at both positions, indicating that thermal processes are the main excitation mechanism of the H$_2$ lines. The resulting excitation temperature are $T_{\rm exc}=2583\pm40$\,K
 and $T_{\rm exc}=2211\pm37$\,K at 0\farcs5 northeast and 0\farcs6 southwest of the nucleus, respectively.

\begin{figure}
\centering
\includegraphics[scale=0.47]{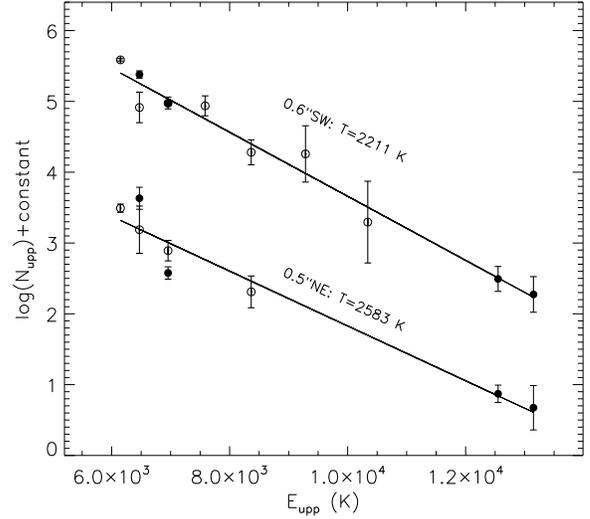}
\caption{Relation between $N_{\rm upp} = \frac{F_i \lambda_i}{A_i g_i}$ and $E_{\rm upp}=T_i$ for the H$_2$ 
emission lines for thermal excitation at the locations of the radio hotspots. {\it Ortho (S)} transitions are shown as filled circles and {\it para (Q)} transitions
 as  open circles.}
\label{excitation}
\end{figure}

The \h2$\lambda$2.1218/\br\ emission-line ratio map shown in Fig.~\ref{ratio} can also be used to investigate the nature of the H$_2$ excitation. 
For Starburst galaxies and H~{\sc ii} regions this ratio is \h2/\br$<$0.6. Larger values are observed for Seyfert nuclei, where the heating of the gas by shocks and X-rays provide additional thermal excitation. Typical values for Seyfert galaxies are in the range 0.6$<$\h2$\lambda$2.1218/\br$<$2.0  \citep{rodriguez-ardila04,rodriguez-ardila05a}, although recent studies suggest a broader range of values 0.4$<$\h2$\lambda$2.1218/\br$<$6.0 for AGNs, including LINERs \citep{rogerio13}. NGC~5929 presents values larger than 2.0 at most locations, confirming that thermal processes are dominating the H$_2$ excitation.  The only exception is a small region with values of $\sim$0.3 at  $\sim$0\farcs8 north-northeast of the nucleus, where the \feii/\pb\ also presents small values. Typical values to the northeast are \h2/\br$\sim$1.0, while to the southwest values of up to 4.5  are observed in regions surrounding the radio knot, suggesting that shocks due to the radio jet contribute to the H$_2$ emission at this location. Similar high values are also observed at the nucleus and along the SE-NW strip, where the interaction of the outflows seen in Paper I with the ambient gas might enhance the H$_2$ emission.  

We also constructed channel maps for the \h2/\br\ ratio in order to investigate the origin of the \h2\ emission at distinct kinematics. These channel maps are shown in Fig.~\ref{chamapH2Br}. The range of velocities is much smaller than those seen for the \feii/\pb\ channel maps in Fig.~\ref{chamapFePb} and the maps are much noisier due to a lower S/N ratio at the K-band lines. At locations of the radio knots, the \h2/\br\ values are larger than 2.0, supporting a the contribution of shocks due to the radio jet to the H$_2$ excitation at these positions, in particular at the southwestern radio knot, where the highest values of \h2/\br\ are observed at the highest velocities channel maps.

\begin{figure*}
\centering
\includegraphics[scale=0.74]{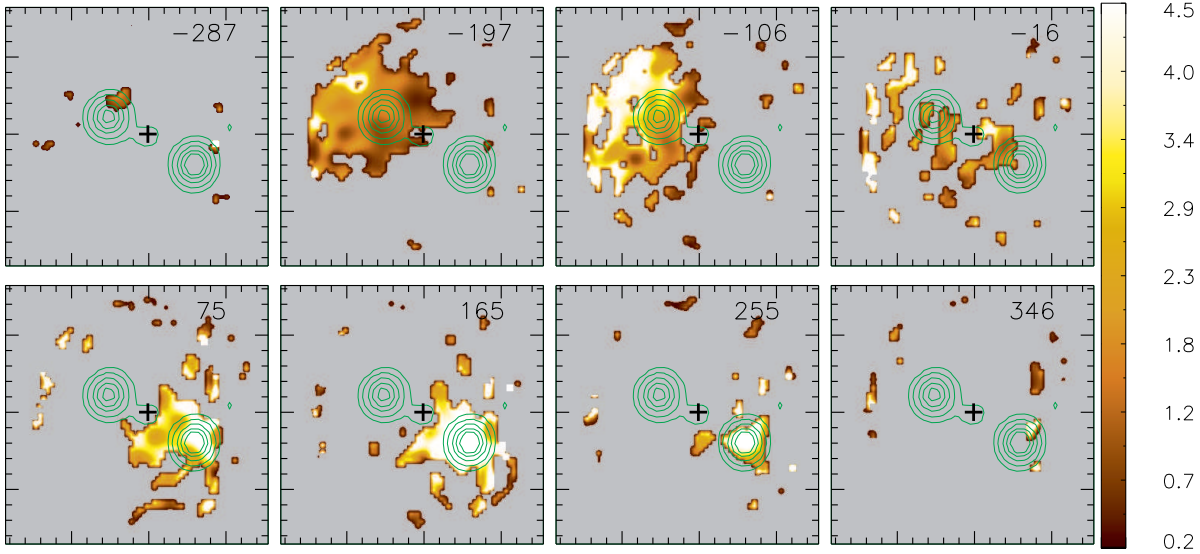}
\caption{Same as Fig.~\ref{chamapFePb} for the \h2$\lambda$2.1218/\br\ line ratio.}
\label{chamapH2Br}
\end{figure*}

Finally, the  \h2$\lambda$2.2477/$\lambda$2.1218 can also be used to distinguish between excitation of the H$_2$ by thermal processes and fluoresce. For thermal processes, this ratio is $\sim0.1-0.2$ and for fluorescent excitation it is $\sim0.55$ \citep[e.g.][]{mouri94,reunanen02,rodriguez-ardila04,sb09}. The \h2$\lambda$2.2477/$\lambda$2.1218 ratio map is shown in Fig.~\ref{ratio} and presents values ranging from 0.1 to 0.25, supporting the conclusion above that fluorescent excitation is not important for NGC~5929.

Thus, we conclude that the H$_2$ emission observed in NGC~5929 is excited by thermal processes, due to heating of the gas by shocks and X-rays from the central AGN. At locations co-spatial with the radio jet and along the SE-NW strip shocks play an important role as indicated by the enhancement of the line ratios. Away from these locations, X-ray heating may dominate the H$_2$ excitation.

\subsection{The origin of the double components}\label{origin-double}

Do the double components observed along the SE-NW strip have distinct origin than the emission from the disk? We can better investigate the origin of the emission of the outflowing gas by the emission-line ratios. Figure~\ref{cut-ratio} 
  presents  the \feii/\pb\ and \h2/\br\ emission-line ratios for the blue (asterisks) and red (diamonds) components along a pseudo slit with 0\farcs35 width oriented along the SE-NW strip at $PA=-30/150^\circ$.  For the \feii/\pb\, both components show typical values of Seyfert galaxies, with average values of 1.0 and 1.3 for the blue and red components, respectively. The \h2/\br\ show some values larger than 2 at some positions, but at most locations these ratio show values typical of Seyfert galaxies. The average values are 1.3 and 2.1 for the blue and red components, respectively. As the line ratio of both components show typical values of Seyfert galaxies, the emission of the outflowing gas might be due to heating by X-rays from the central AGN.

\begin{figure}
\centering
\includegraphics[scale=0.55]{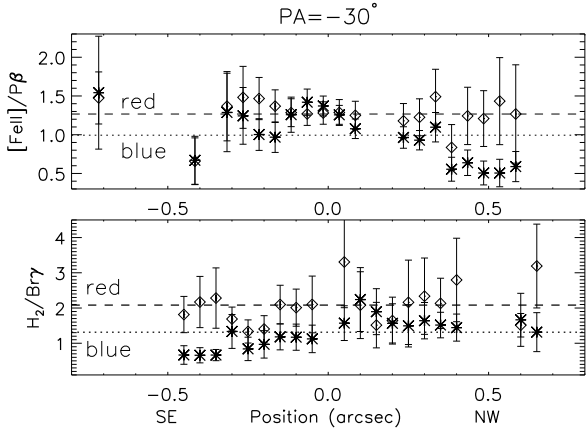}
\caption{\feii/\pb\ and \h2/\br\ emission-line ratios along $PA=-30/150^\circ$. The red component is shown as open diamonds and the blue component as asterisks. The dashed line shows the average value for the red component and the dotted line represents the average ratio for the blue component.}
\label{cut-ratio}
\end{figure}

\section{Conclusions}

We used Gemini integral field J- and K$_{\rm l}$-band spectroscopy of the inner 250\,pc radius of the Seyfert 2 galaxy NGC\,5929,  at spatial resolution 20\,pc and velocity resolution 40~\kms, to map the stellar and gas kinematics and the flux distributions of emission lines  from the ionized and warm molecular gas.  Our main conclusions are:

\begin{itemize}

\item The flux distributions for the \h2, H recombination and forbidden emission lines are extended to up to 250~pc, with the highest flux levels oriented along $PA=60/240^\circ$ (the direction of the radio jet), and well correlated with the radio emission. The \h2\ emission is more distributed over the whole field of view, while the emission of the ionized gas is more collimated along the direction of the radio jet.

\item The excitation of the \h2\ and \feii\ at all locations is dominated by thermal processes, mainly due to heating of the gas by X-rays emitted emitted by the central AGN. Some contribution from shocks is observed in the high velocity gas at locations next to the radio hotpots, as indicated by the \feii/\pb\ and \h2/\br\ line-ratio channel maps.

\item The stellar velocity field shows rotation with an amplitude of up to $\sim$200\,\kms\ when corrected for the inferred inclination of 18.3$^\circ$, and a velocity dispersion reaching 180\,\kms. From  the  $M_{\bullet}-\sigma$ relationship, we estimated a mass for the supermassive black hole of  $M_\bullet=5.2^{1.6}_{-1.2}\times10^7~{\rm M_\odot}$ using the median stellar velocity dispersion of $\sigma_*=133\pm8$~\kms.

\item The gas kinematics present three components: (i) a counter-rotating disk (relative to the stellar velocity field); (ii) an equatorial outflow (perpendicular to the radio jet), thus in the plane of the torus, with a mass-outflow rate of $\dot{M} > 0.4~{\rm M_\odot\, yr^{-1}}$ and (iii) kinematic disturbances observed in association with the radio hot spots in blueshift to the northeast and redshift to the southwest, what supports that the radio jet is tilted towards us at the northeast and away from us at the southwest.
  
\item  From the  \br\  and \h2\ emission-line fluxes, we calculate the mass of ionized and warm molecular gas of $M_{HII}=1.3\pm0.2\times10^6\,{\rm M_\odot}$
  and $M_{H_2}=471\pm58~{\rm M_\odot}$, respectively. These values correspond to mass surface densities of  
$\sum_{M_{H_2}}=1.7\times10^{-3}{\rm M_\odot/pc^2}$ and $\sum_{M_{HII}}=4.8 {\rm M_\odot/pc^2}$, which are in the range 
of values observed for other galaxies.

\end{itemize}

Our favored scenario for this galaxy is that the interaction with NGC\,5930 has sent gas towards the nucleus of NGC5929, triggering the nuclear activity. If the observed equatorial outflow can be considered the first blast of the AGN, we estimate that this happened less than 1\,Myr ago.

\section*{Acknowledgments}
 We thank the referee for his/her thorough review, comments and
suggestions, which helped us to significantly improve this paper.
Based on observations obtained at the Gemini Observatory, 
which is operated by the Association of Universities for Research in Astronomy, Inc., under a cooperative agreement with the 
NSF on behalf of the Gemini partnership: the National Science Foundation (United States), the Science and Technology 
Facilities Council (United Kingdom), the National Research Council (Canada), CONICYT (Chile), the Australian Research 
Council (Australia), Minist\'erio da Ci\^encia e Tecnologia (Brazil) and south-eastCYT (Argentina).  
This research has made use of the NASA/IPAC Extragalactic Database (NED) which is operated by the Jet
 Propulsion Laboratory, California Institute of  Technology, under contract with the National Aeronautics and Space Administration.
{\it R.A.R.} acknowledges support from FAPERGS (project N0. 2366-2551/14-0) and CNPq (project N0. 470090/2013-8 and 302683/2013-5).

\label{lastpage}

\end{document}